\documentclass[12pt,a4paper]{article}
\usepackage{a4wide}
\usepackage{amsmath}
\usepackage{amssymb}
\usepackage{epsfig}
\usepackage{subfigure}
\usepackage{exscale}
\usepackage{float}
\usepackage{bbm}
\usepackage[numbers,sort&compress]{natbib}
\usepackage{pst-plot, pstricks,pst-math}
\usepackage{fancybox,amssymb,color}
\usepackage{graphicx}
\usepackage{pstricks, color, graphicx, epsfig, psfrag}
\usepackage{amsfonts,amsmath,amssymb,slashed}
\usepackage{dsfont}
\usepackage{bbm,bm}
\usepackage{fancyhdr, a4wide}
\usepackage[english]{babel}
\usepackage{subfigure}

\makeatletter
\@addtoreset{equation}{section}
\makeatother

\title{Thermodynamics of Two-Dimensional Ideal Ferromagnets -- Three-Loop Analysis}

\author{Christoph P.\ Hofmann$^a$ \\ \\
\normalsize {$^a$ Facultad de Ciencias, Universidad de Colima} \\
\vspace{0.3cm}
\normalsize {Bernal D\'iaz del Castillo 340, Colima C.P.\ 28045, Mexico} \\}

\begin{document}

\maketitle

\begin{abstract} \normalsize

\end{abstract}

Within the effective Lagrangian framework, we explicitly evaluate the partition function of two-dimensional ideal ferromagnets up to
three loops at low temperatures and in the presence of a weak external magnetic field. The low-temperature series for the free energy
density, energy density, heat capacity, entropy density, as well as the magnetization are given and their range of validity is critically
examined in view of the Mermin-Wagner theorem. The calculation involves the renormalization and numerical evaluation of a particular
three-loop graph which is discussed in detail. Interestingly, in the low-temperature series for the two-dimensional ideal ferromagnet, the
spin-wave interaction manifests itself in the form of logarithmic terms. In the free energy density the leading such term is of order
$T^4 \ln T$ -- remarkably, in the case of the three-dimensional ideal ferromagnet no logarithmic terms arise in the low-temperature
series. While the present study demonstrates that it is straightforward to consider effects up to three-loop order in the effective field
theory framework, this precision seems to be far beyond the reach of microscopic methods such as modified spin-wave theory.

% \pacs{12.39.Fe, 75.30.Ds, 11.10.Wx}

\maketitle

\section{Introduction}
\label{Intro}

In a very recent article on the thermodynamic properties of two-dimensional ideal ferromagnets \citep{Hof12}, the general structure of the
low-temperature series for the free energy density has been discussed using the method of effective Lagrangians. In particular, it has
been argued that the spin-wave interaction does not yet manifest itself at order $T^3$ in the free energy density -- it only shows up at
order $T^4 \ln T$. The explicit evaluation of the various Feynman graphs contributing at this order, however, has not been performed in
that reference -- this will be the subject of the present article. The calculation involves a particular three-loop graph whose
renormalization and subsequent numerical evaluation -- although quite elaborate -- nevertheless is rather straightforward within the
effective field theory framework. While Ref.~\citep{Hof12} focused on the low-temperature series for the free energy density, here we also
consider the magnetization and critically examine the range of validity of the corresponding low-temperature series -- indeed, in view of
the Mermin-Wagner theorem, one has to be very careful by taking the limit of a zero external magnetic field $H$.

Interestingly, these logarithmic contributions  $T^n \ln T$ in the partition function are restricted to two-dimensional ferromagnets and
do not show up in the case of three-dimensional ideal ferromagnets. There the low-temperature series of the various thermodynamic
quantities consist of integer and half-integer powers of the temperature \citep{Dys56,Zit65,Hof02,Hof11a}. Actually, the occurrence of such
logarithmic contributions is well-known in the domain of particle physics -- the low-temperature expansion of e.g. the free energy density
in Quantum Chromodynamics also involves a logarithmic term $T^8 \ln T$ \citep{GL89}. A term $T^8 \ln T$ also occurs in the low-temperature
expansion of the free energy density of the three-dimensional antiferromagnet \citep{Hof99b}, which obeys a linear, i.e. relativistic,
dispersion law.

We also confront our series obtained within the effective Lagrangian framework with the relevant literature and point out that our
approach is much more efficient than conventional condensed matter methods such as spin-wave theory or Schwinger-Boson mean field theory.
In particular, while all these different studies were restricted to the idealized picture of the free magnon gas, in the present article
we discuss in detail how the spin-wave interaction manifests itself in the low-temperature behavior of the two-dimensional ideal
ferromagnet.

The rest of the paper is organized as follows. In Sec.~\ref{EFT} we briefly discuss some essential aspects of the effective Lagrangian
method -- detailed accounts can be found in the references provided below. The evaluation of the partition function up
to three-loop order in the low-temperature expansion is presented in Sec.~\ref{PartitionFunction}. While the renormalization up to order
$p^6 \propto T^3$ turns out to be straightforward, the handling of ultraviolet divergences at order $p^8$ is more involved and is
considered in detail in Sec.~\ref{Renorm}. The low-temperature expansions for the free energy density, energy density, heat
capacity, and entropy density of the two-dimensional ideal ferromagnet are given in Sec.~\ref{Thermodyn}. The low-temperature series for
the magnetization is discussed in Sec.~\ref{Magnetization} and the range of validity of this series is critically examined in view of the
Mermin-Wagner theorem. Here we also compare our results with the condensed matter literature. While our conclusions are presented in
Sec.~\ref{Summary}, the numerical evaluation of a specific three-loop graph, the logarithmic renormalization of effective constants and
the representation for the pressure at zero magnetic field are relegated to three separate appendices.

Unfortunately, the systematic and model-independent effective Lagrangian method is still not very well known within the condensed matter
community. Therefore we would like to provide the reader with a brief list of references where condensed matter problems have been
successfully solved within the effective field theory framework. These include antiferromagnets and ferromagnets in two and three spatial
dimensions \citep{Hof02,Hof11a,HL90,HN91,HN93,Leu94a,Hof99a,Hof99b,RS99a,RS99b,RS00,Hof01,Hof10,Hof11b,Hof11c},
as well as two-dimensional antiferromagnets which are the precursors of high-temperature superconductors
\citep{KMW05,BKMPW06,BKPW06,BHKPW07,BHKMPW07,BHKMPW09,JKHW09,KBWHJW12,VHJW12,JKBWHW12}.

Many different tests -- both analytical and numerical studies -- unambiguously confirm that the effective Lagrangian technique represents
a rigorous and systematic framework to address condensed matter systems which exhibit a spontaneously broken symmetry. On the one hand,
the correctness of the effective Lagrangian approach was demonstrated explicitly in Ref.~\citep{GHKW10} by comparing the microscopic
results of an analytically solvable model for a hole-doped ferromagnet in 1+1 dimensions with the effective theory predictions. On the
other hand, in a series of high-accuracy investigations of the antiferromagnetic spin-$\frac{1}{2}$ quantum Heisenberg model on a square
lattice using the loop-cluster algorithm \citep{WJ94,GHJNW09,JW11,GHJPSW11}, the Monte Carlo data were confronted with the analytic
predictions of the effective Lagrangian theory and the low-energy constants were extracted with permille accuracy.

\section{Essential Aspects of Effective Lagrangians}
\label{EFT}

The present article deals with the explicit evaluation of the partition function of the two-dimensional ideal ferromagnet at three-loop
order and the subsequent discussion of various thermodynamic quantities -- including the magnetization -- at low temperatures. It is an
extension of the work presented in the recent article \citep{Hof12}, where the evaluation was restricted to two-loop order. The basic
principles of the effective Lagrangian method and the perturbative evaluation of the partition function have been outlined in section 2 of
that reference and will not be repeated here in detail. The interested reader may also find a detailed account on finite-temperature
effective Lagrangians in appendix A of Ref.~\citep{Hof11a} and in the various references given therein. In addition, for pedagogic
introductions to the effective Lagrangian technique, we refer to Refs.~\citep{Bur07,Brau10,Leu95,Sch03,Goi04}. In this section, we will
just focus on some essential aspects of the effective Lagrangian method at finite temperature.

The effective Lagrangian, or more precisely, the effective action
\begin{equation}
{\cal S}_{eff} = \int d^3 x \, {\cal L}_{eff}
\end{equation}
of the two-dimensional ideal ferromagnet must share all the symmetries of the underlying Heisenberg model, i.e. the spontaneously broken
spin rotation symmetry O(3), parity and time reversal. The basic degrees of freedom of the effective Lagrangian are the two real magnon
fields -- or the physical magnon particle -- which emerge due to the spontaneously broken spin symmetry O(3) $\to$ O(2).

\begin{figure}
\includegraphics[width=13.5cm]{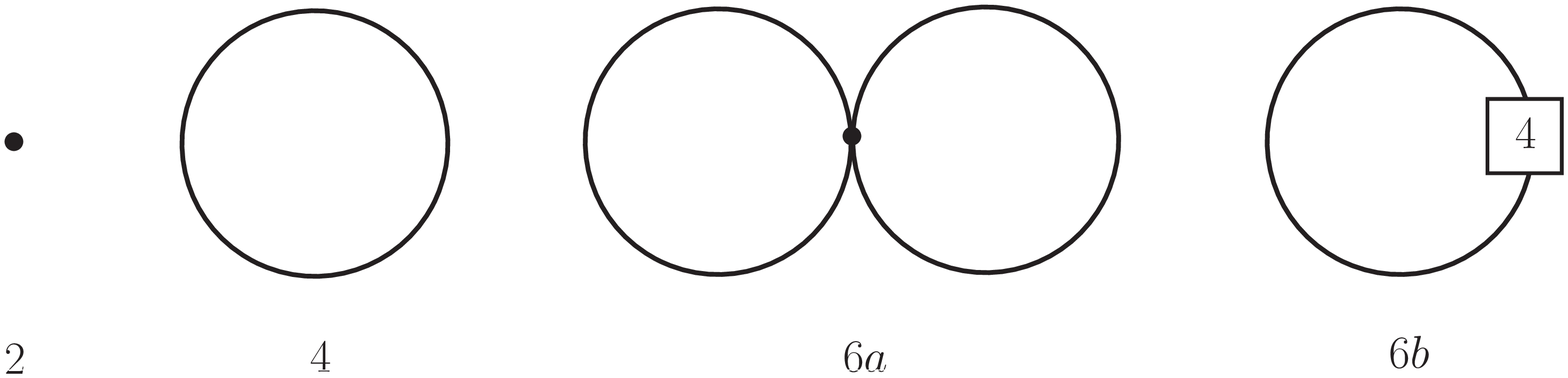}

\vspace{4mm}

\includegraphics[width=13.0cm]{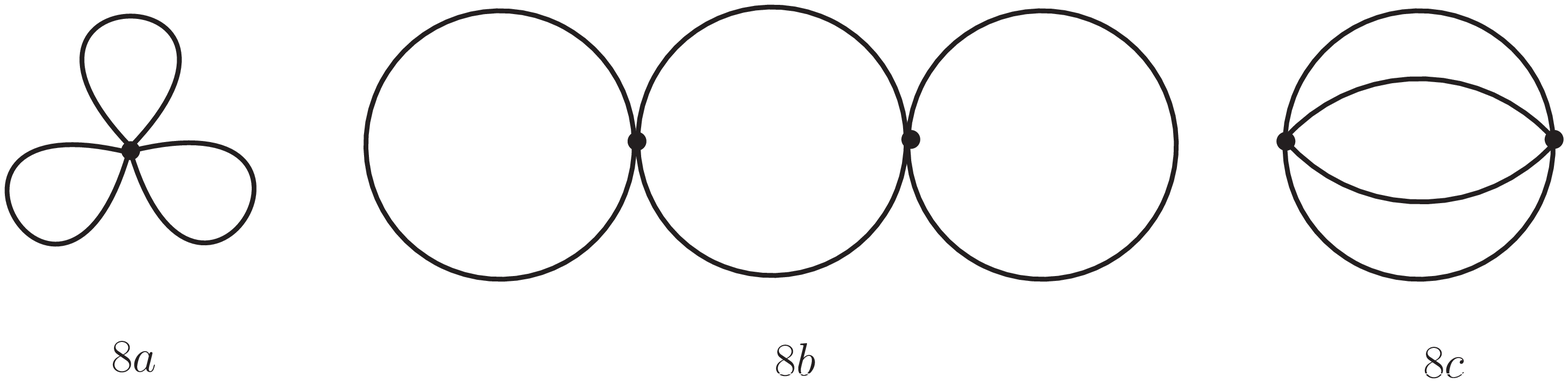}

\vspace{4mm}

\includegraphics[width=10.5cm]{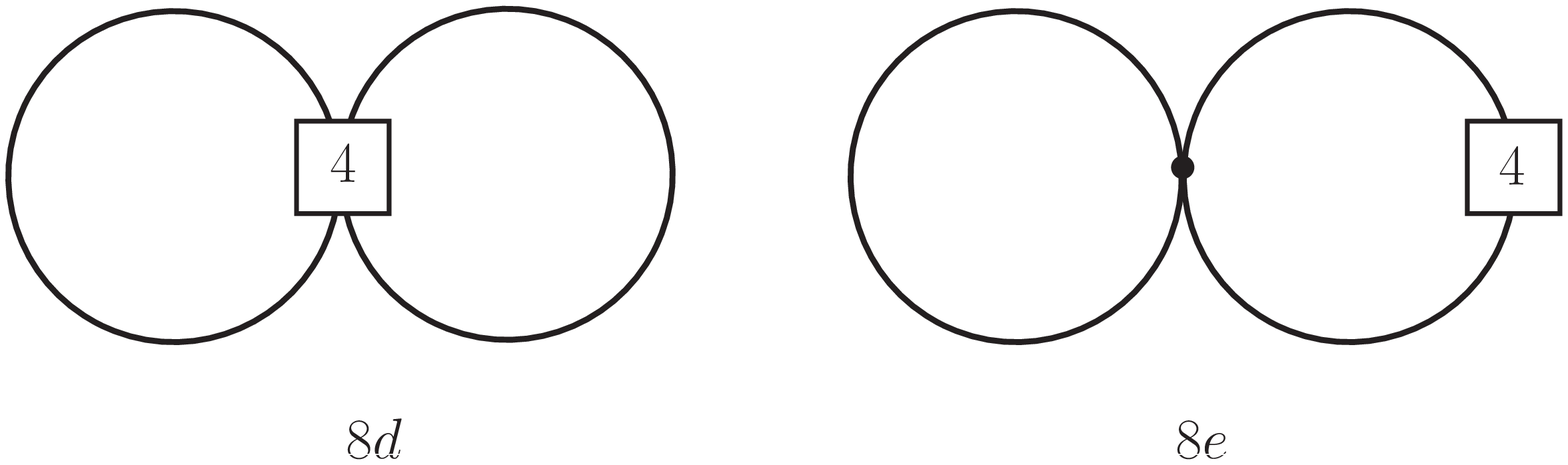}

\vspace{4mm}

\includegraphics[width=7.0cm]{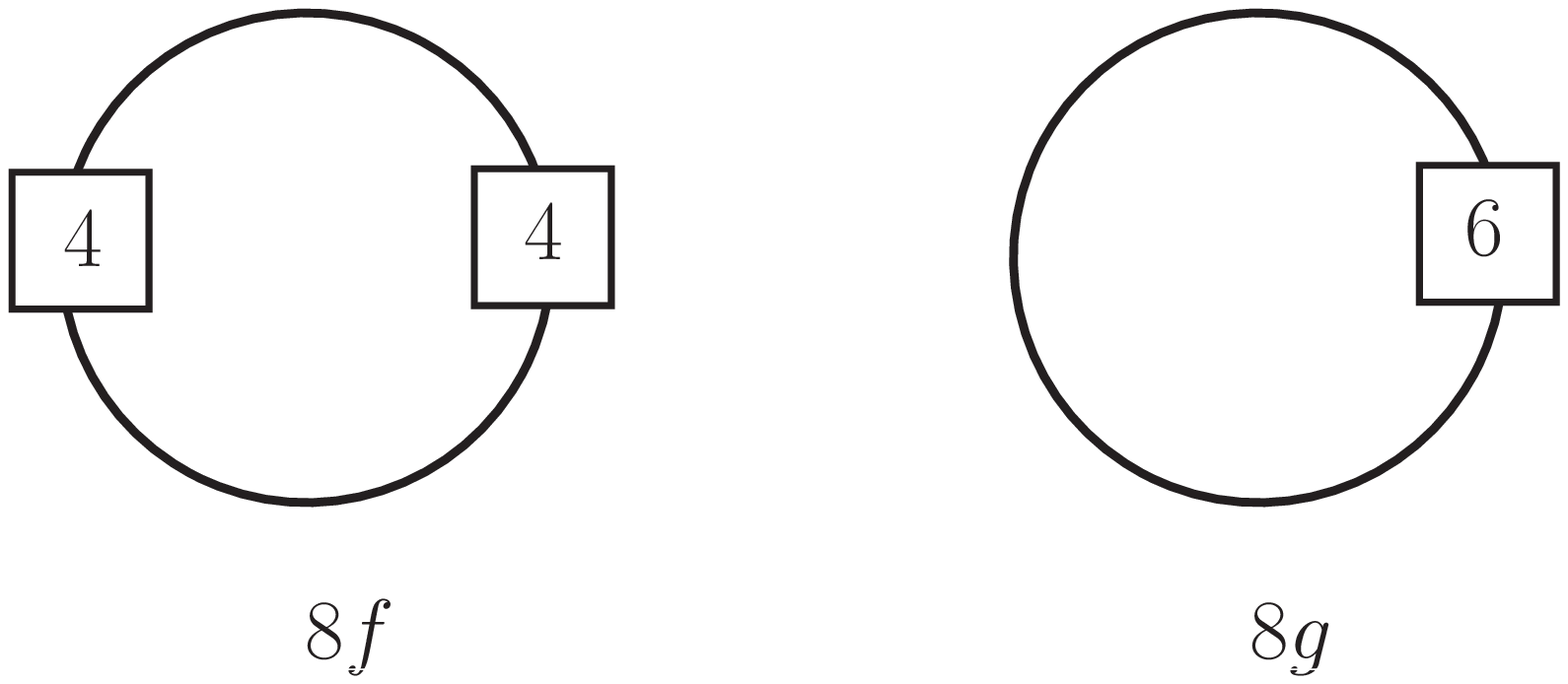}

\caption{Feynman graphs related to the low-temperature expansion of the partition function for a two-dimensional ferromagnet up to order
$p^8$. The numbers attached to the vertices refer to the piece of the effective Lagrangian they come from. Vertices associated with the
leading term ${\cal L}^2_{eff}$ are denoted by a dot. Note that ferromagnetic loops are suppressed by two powers of momentum in two
spatial dimensions, $d_s$=2.}
\label{figure1}
\end{figure} 

The various pieces in the effective Lagrangian are organized according to the number of space and time derivatives which act on the
magnon fields. This derivative expansion -- or expansion in powers of momentum --  is completely systematic: At low energies, terms in the
effective Lagrangian which contain only a few derivatives are the dominant ones, while terms with more derivatives are suppressed
\citep{Wei79,GL85,Leu94b}. The leading-order effective Lagrangian for the ideal ferromagnet is of momentum order $p^2$ and takes the form
\citep{Leu94a}
\begin{equation}
\label{leadingLagrangian}
{\cal L}^2_{eff} = \Sigma \frac{\epsilon_{ab} {\partial}_0 U^a U^b}{1+ U^3}
+ \Sigma \mu H U^3 - \frac{1}{2} F^2 {\partial}_r U^i {\partial}_r U^i \, .
\end{equation}
The two real components of the magnon field, $U^a \, (a=1,2)$ are the first two components of the three-dimensional unit vector
$U^i = (U^a, U^3)$. The quantity $H$ is the third component of the magnetic field ${\vec H} = (0,0,H)$. While the derivative structure of
the above terms is unambiguously determined by the symmetries of the underlying theory, the two a priori unknown low-energy constants --
the spontaneous magnetization at zero temperature $\Sigma$, and the constant $F$ -- have to be determined by experiment, numerical
simulation or comparison with the microscopic theory. Note that one time derivative (${\partial}_0$) is on the same footing as two space
derivatives (${\partial}_r {\partial}_r$), i.e., two powers of momentum count as only one power of energy or temperature:
$k^2 \propto \omega, T$.

The next-to-leading order Lagrangian for the ideal ferromagnet is of order $p^4$ and amounts to \citep{Hof02}
\begin{equation}
\label{Leff4}
{\cal L}^4_{eff} = l_1 {( {\partial}_r U^i {\partial}_r U^i )}^2 +
l_2 {( {\partial}_r U^i {\partial}_s U^i )}^2 + l_3 \Delta U^i \Delta U^i \, ,
\end{equation}
where $\Delta$ denotes the Laplace operator in two spatial dimensions, $d_s=2$, and the quantities $l_1, l_2$ and $l_3$ are effective
coupling constants.

A crucial point underlying the perturbative evaluation of the partition function concerns the suppression of loop diagrams in the
effective field theory framework. In two spatial dimensions, ferromagnetic loops are suppressed by two powers of momentum. This
suppression rule lies at the heart of the organization of the Feynman graphs of the partition function for the two-dimensional ferromagnet
depicted in Fig.~\ref{figure1}. For example, the two-loop diagram 8d with an insertion from ${\cal L}^4_{eff}$, containing four magnon
fields, is of order $p^8$, as it involves ${\cal L}^4_{eff}$ ($p^4$) and two loops ($p^4$).

An inspection of the diagrams in Fig.~\ref{figure1} reveals that insertions from ${\cal L}^6_{eff}$ only appear in the one-loop graph 8g.
The only term we need for the present evaluation is thus quadratic in the magnon field,
\begin{equation}
\label{Leff6}
{\cal L}^6_{eff} = c_1 U^i {\Delta}^3 U^i \, ,
\end{equation}
where $c_1$ is an additional effective low-energy constant.

Lattice anisotropies do not yet manifest themselves in the leading-order effective Lagrangian ${\cal L}^2_{eff}$ -- the expression
(\ref{leadingLagrangian}) is invariant under continuous space rotations, even though the underlying lattices (square, honeycomb,
triangular and Kagom\'e, to name the most prominent ones) are only invariant under discrete space rotations. Although this accidental
symmetry is restricted to ${\cal L}^2_{eff}$, in the present analysis we assume that the higher-order pieces ${\cal L}^4_{eff}$ and
${\cal L}^6_{eff}$ are O(3)-space-rotation invariant as well. The conclusions of the present paper will not be affected by this
idealization.

\section{Partition Function up to Three Loops}
\label{PartitionFunction}

While the evaluation of the partition function of the two-dimensional ideal ferromagnet up to order $p^6$ was presented in
Ref.~\citep{Hof12}, here we go one step beyond and consider the evaluation up to order $p^8$, where three-loop graphs start to show up.
This calculation is rather involved due to the renormalization and numerical evaluation of the three-loop graph 8c.

As depicted in Fig.~\ref{figure1}, there are only four Feynman diagrams that contribute to the partition function up to order $p^6$ or,
equivalently, up to order $T^3$. The evaluation of these diagrams can be found in Ref.~\citep{Hof12} -- the final expression for the free
energy density amounts to
\begin{equation}
\label{FreeCollect}
z \; = \; - \Sigma \mu H \; - \; \frac{1}{4 \pi \gamma} \, T^2 \, \sum^{\infty}_{n=1} \frac{e^{- \mu H n \beta}}{n^2}
\; - \; \frac{l_3}{\pi \Sigma {\gamma}^3} \, T^3 \, \sum^{\infty}_{n=1} \frac{e^{- \mu H n \beta}}{n^3} \; + \; {\cal O}(p^8) \, ,
\end{equation}
where $\beta \equiv 1/T$ and $H$ is the external magnetic field. The terms of order $T^2$ and $T^3$ arise from one-loop graphs and thus
represent contributions to the free energy density originating from noninteracting spin waves. While the former term is exclusively
determined by the leading-order effective constants $\Sigma$ and $F$ ($\gamma = F^2/\Sigma$), the term of order $T^3$, on the other hand,
involves the next-to-leading-order effective constant $l_3$ from ${\cal L}^4_{eff}$. What is quite remarkable is that the spin-wave
interaction does not yet manifest itself at this order in the low-temperature expansion of the free energy density. The only potential
candidate, the two-loop diagram 6a of order $T^3$, turns out to be zero due to parity \citep{Hof12}.

Before we evaluate the partition function at order  $p^8$, we have to recall that, at finite temperature, the propagator is given by
\begin{equation}
\label{ThermalPropagator}
G(x) \, = \, \sum_{n \,= \, - \infty}^{\infty} \Delta({\vec x}, x_4 + n \beta) \, ,
\end{equation}
where $\Delta(x)$ is the Euclidean propagator of ferromagnetic magnons at zero temperature,
\begin{equation}
\label{Propagator}
\Delta (x) \, = \, \int \! \, \frac{d k_4 d^2\!k}{(2\pi)^3}
\frac{e^{i{\vec k}{\vec x} - i k_4 x_4}}{\gamma {\vec k}^2 - i k_4 + \mu H} \, . 
\end{equation}
An explicit representation for the thermal propagator, dimensionally regularized in the spatial dimension $d_s$, is
\begin{equation}
\label{ThermProp}
G(x) = \frac{1}{(2{\pi})^{d_s}} \, \Big(\frac{{\pi}}{\gamma}\Big)^{\frac{d_s}{2}}
\sum^{\infty}_{n \, = \, - \infty} \frac{1}{x_n^{\frac{d_s}{2}}} \,
\exp \Big[ - \frac{{\vec x}^2}{4 \gamma x_n} - \mu H x_n \Big] \,
\Theta (x_n) \, ,
\end{equation}
with
\begin{equation}
x_n \, \equiv \, x_4 + n \beta \, .
\end{equation}

Also, it is convenient to introduce the following notation,
\begin{equation}
\label{definitionsThermProp}
G_1 \equiv \Big[ G(x) \Big]_{x=0} \, , \quad
G_{\Delta} \equiv \Big[ {\Delta} G(x) \Big]_{x=0} \, , \quad
G_{\Delta^n} \equiv \Big[ {\Delta}^n G(x) \Big]_{x=0} \, ,
\end{equation}
where $\Delta$ represents the Laplace operator in the spatial dimensions -- note that the quantity $\Delta (x)$, on the other hand, stands
for the zero-temperature propagator.

The quantities $G_1$, $G_{\Delta}$, as well as thermal propagators involving even higher-order space derivatives, can be split into a
finite piece, which is temperature dependent, and a divergent piece, which is temperature independent,
\begin{equation}
\label{DecompositionPropagator}
G_1 \; = \; G^T_1 \, + \, G^0_1 \, , \qquad G_{\Delta} \; = \; G^T_{\Delta} \, + \, G^0_{\Delta} \, .
\end{equation}
The explicit dimensionally regularized expressions are
\begin{eqnarray}
\label{ThermProp(x=0)}
G^T_1 & = & \frac{1}{(2{\pi})^{d_s}} \,
\Big(\frac{\pi}{\gamma}\Big)^{\frac{d_s}{2}} \, \sum^{\infty}_{n=1}
\frac{e^{- \mu H n \beta}}{(n \beta)^{\frac{d_s}{2}}} \, , \nonumber \\
G^0_1 & = & \frac{1}{(2{\pi})^{d_s}} \, \Big(\frac{{\pi}}{\gamma}\Big)^{\frac{d_s}{2}}
\Bigg [\frac{1}{{x_4}^{\frac{d_s}{2}}} \exp\Big(-\frac{{\vec x}^2}{4 \gamma x_4} \Big)
\Theta(x_4) \Bigg]_{x=0} \, ,
\end{eqnarray}
and
\begin{eqnarray}
\label{DeltaProp(x=0)}
G^T_{\Delta} & = & \frac{1}{(2{\pi})^{d_s}} \, \Big(\frac{\pi}{\gamma}\Big)^{\frac{d_s}{2}} \,
\Big(\! - \frac{d_s}{2\gamma} \Big) \sum^{\infty}_{n=1} \frac{e^{ - \, \mu H n
\beta}}{(n\beta)^{\frac{d_s}{2} + 1}} \, , \nonumber \\
G^0_{\Delta} & = & \frac{1}{(2{\pi})^{d_s}} \, \Big(\frac{\pi}{\gamma}\Big)^{\frac{d_s}{2}}
\Bigg[ \frac{1}{{x_4}^{\frac{d_s}{2} + 1}} \Bigg\{ \frac{-d_s}{2 \gamma} + \frac{{\vec x}^2}
{4 {\gamma}^2 x_4} \Bigg\} \exp\Big(-\frac{{\vec x}^2}{4 \gamma x_4} \Big)
\Theta(x_4) \Bigg]_{x=0} \, .
\end{eqnarray}
The temperature-independent pieces $G^0_1, G^0_{\Delta}, \dots$ are all related to momentum integrals of the form
\begin{equation}
\label{RuleDimReg}
\int \! \, d^{d_s}\!k \, \Big( {\vec k}^2 \Big)^m
\exp\Big[ - \gamma x_4 {\vec k}^2 - x_4 \mu H \Big] \, , \qquad m = 0, 1, 2,
\ldots \, ,
\end{equation}
which are proportional to
\begin{equation}
\label{RuleDimReg2}
\frac{\exp[- x_4 \mu H]}{(\gamma x_4)^{m+\frac{d_s}{2}}} \, \,
\Gamma\Big(m+\frac{d_s}{2}\Big) \, .
\end{equation}
In dimensional regularization these expressions vanish altogether: $G^0_1, G^0_{\Delta}$, and zero-temperature propagators involving
higher-order space derivatives do not contribute in the limit $d_s \! \to \! 2$. The only contributions we are thus left with here, are
those which involve the temperature-dependent pieces.

According to Fig.~\ref{figure1}, there are seven diagrams at order $p^8$ we need to evaluate. We first consider the two one-loop graphs
which contain vertices from ${\cal L}^4_{eff}$ and ${\cal L}^6_{eff}$. For graph 8g, which only involves an insertion from
${\cal L}^6_{eff}$, we obtain
\begin{equation}
\label{z(8g)}
z_{8g} \ = \ - \frac{2 \, c_1}{\Sigma} \, G_{\Delta^3} \, ,
\end{equation}
yielding the temperature-dependent contribution
\begin{equation}
\label{z(8g)T}
z^T_{8g} \ = \ \frac{3 \, c_1}{\pi \Sigma {\gamma}^4} \, T^4 \, \sum^{\infty}_{n=1} \frac{e^{- \mu H n \beta}}{n^4} \, \, .
\end{equation}
Graph 8f, which contains two insertions from ${\cal L}^4_{eff}$, is proportional to an integral over the torus
${\cal T} = {\cal R}^{d_s} \times S^1$, with circle $S^1$ defined by $- \beta / 2 \leq x_4 \leq \beta / 2$, and involves a product of two
thermal propagators,
\begin{equation}
\label{z(11e)}
z_{8f} \, = \, - \frac{2 l^2_3}{{\Sigma}^2} \, {\int}_{\! \! \! {\cal T}} \! \! d^{d_s+1}x \, {\Delta}^2 G(x) \, {\Delta}^2 G(-x) \, .
\end{equation}
Integrals of this type, as shown in Ref.~\citep{Hof02}, can be converted into an expression displaying one propagator only,
\begin{equation}
\label{2->1Delta4}
- {\int}_{\! \! \! {\cal T}} \! \! d^{d_s+1}y \, {\Delta}^m G(-y) \, {\Delta}^n G(y) \,  = \, \Bigg[ {\Delta}^{(m+n)}
\frac{\partial G(x)}{\partial (\mu H)} \Bigg]_{x=0} \, ,
\end{equation}
such that one ends up with
\begin{equation}
z_{8f} \, = \, \frac{2 l^2_3}{{\Sigma}^2} \, \Bigg[ {\Delta}^4  \frac{\partial G(x)}{\partial (\mu H)} \Bigg]_{x=0} \, .
\end{equation}
Accordingly, the temperature-dependent part of graph 8f amounts to
\begin{equation}
\label{z(8f)T}
z^T_{8f} \ = \ - \frac{12 l_3^2}{\pi {\Sigma}^2 {\gamma}^5} \, T^4 \, \sum^{\infty}_{n=1} \frac{e^{- \mu H n \beta}}{n^4} \, \, .
\end{equation}

We now turn to the two-loop graphs which involve insertions from ${\cal L}^2_{eff}$ and ${\cal L}^4_{eff}$. Graph 8d contributes with
\begin{equation}
\label{z(8d)}
z_{8d} = - \frac{2}{3 {\Sigma}^2} (8 l_1 + 6 l_2 + 5 l_3) \, G_{\Delta} G_{\Delta} \, - \frac{2 l_3}{{\Sigma}^2} \, G_1  G_{\Delta^2} \, .
\end{equation}
The evaluation of graph 8e yields
\begin{equation}
\label{z(8e)}
z_{8e} = \frac{2 l_3}{{\Sigma}^2} G_1  G_{\Delta^2} \, ,
\end{equation}
and cancels the second term in $z_{8d}$. For the temperature-dependent part originating from the two-loop graphs of order $p^8$ we thus
end up with
\begin{equation}
\label{z(8de)}
z^T_{8[de]} = - \frac{8 l_1 + 6 l_2 + 5 l_3}{24 {\pi}^2 {\Sigma}^2 {\gamma}^4} \, T^4 \, {\Bigg\{ \sum^{\infty}_{n=1}
\frac{e^{- \mu H n \beta}}{n^2} \Bigg\}}^2 \, \, .
\end{equation}

Finally, we consider the three-loop graphs. Note that they exclusively contain vertices from the leading-order Lagrangian
${\cal L}^2_{eff}$. Graph 8a factorizes into a product of three thermal propagators (and space derivatives thereof), to be evaluated at
the origin,
\begin{equation}
z_{8a} \, = \, - \frac{F^2}{{\Sigma}^3} \, G_{\Delta} {(G_1)}^2 \, .
\end{equation}
Remarkably, the three-loop graph 8b turns out to be zero,
\begin{equation}
z_{8b} = 0 \, .
\end{equation}
Much like the two-loop graph 6a, the three-loop graph 8b does not contribute to the partition function.

We are left with the cateye graph 8c which leads to
\begin{equation}
\label{cateye}
z_{8c} \, = \, - \frac{F^4}{2{\Sigma}^4} \, K \, + \, \frac{F^2}{{\Sigma}^3 }
G_{\Delta} {(G_1)}^2 \, .
\end{equation}
The quantity $K$ denotes the following integral over the torus involving a product of four thermal propagators
\begin{equation}
\label{integralK}
K \, = {\int}_{\! \! \! {\cal T}} \! \! d^{d_s+1}x \, {\partial}_r G \,
{\partial}_r G \, {\partial}_s {\tilde G} \, {\partial}_s {\tilde G} \, .
\end{equation}
Here we have used the notation
\begin{equation}
G = G(x) \, , \qquad {\tilde G} = G(-x) \, .
\end{equation}
The second term in (\ref{cateye}) cancels the contribution from graph 8a -- hence the overall contribution from the three-loop graphs is
just the one proportional to the integral $K$. Note that, unlike all other pieces in the free energy density up to order $p^8$, this
quantity is not a product of thermal propagators (or derivatives thereof) to be evaluated at the origin -- its structure is much more
complicated. In the next section, as well as in the appendices \ref{AppendixA} and \ref{AppendixB}, we will address in detail the
renormalization and numerical evaluation of this integral which contains a total of four infinite sums.

\section{Renormalization of the Cateye Graph}
\label{Renorm}

In order to renormalize the three-loop graph 8c, we will follow the method outlined in Ref.~\citep{GL89}, where the same graph was
considered in a Lorentz-invariant framework.

The relevant integral from the three-loop contribution is given by
\begin{displaymath}
K \, = {\int}_{\! \! \! {\cal T}} \! \! d^{d_s+1}x \, {\partial}_r G \,
{\partial}_r G \, {\partial}_s {\tilde G} \, {\partial}_s {\tilde G},
\end{displaymath}
and involves a product of four thermal propagators. In order to analyze the limit $d_s \! \to \! 2$, the thermal propagators are split
into two pieces,
\begin{equation}
G(x) = G^T(x) + \Delta(x) \, .
\end{equation}
While the zero-temperature propagator $\Delta(x)$ contains the ultraviolet singularities, the temperature-dependent part $G^T(x)$ is
finite as $d_s \! \to \! 2$. Note that, if we restrict ourselves to the origin, we reproduce the first relation of
Eq.(\ref{DecompositionPropagator}).

With the above decomposition, the integral $K$ yields nine terms that can be grouped into the following six classes -- for simplicity we
do not display the derivatives:
\begin{eqnarray}
\label{SixTypes}
A:& & G^T(x) \, G^T(x) \, G^T(-x) \, G^T(-x) ,
\nonumber \\
B:& & \Delta(x) \, G^T(x) \, G^T(-x) \, G^T(-x) \, , \ G^T(x) \, G^T(x) \, \Delta(-x) \, G^T(-x) , \nonumber \\
C:& & {\Delta}^2(x) \, G^T(-x) \, G^T(-x) \, , \
G^T(x) \, G^T(x)  \,\Delta^2(-x), \nonumber \\
D:& & \Delta(x) \, G^T(x) \, \Delta(-x) \, G^T(-x)\, ,
\nonumber \\
E:& & {\Delta}^2(x) \, \Delta(-x) \, G^T(-x) \, , \
\Delta(x) \, G^T(x) \, {\Delta}^2(-x) \, , \nonumber \\
F:& & {\Delta}^2(x) \, {\Delta}^2(-x).
\end{eqnarray}
As the product $\Delta(x) \Delta(-x)$ of zero-temperature propagators involves the combination $\Theta(x_4) \Theta(-x_4)$, terms of the
classes $D, E$ and $F$ vanish identically. The maximum number of $\Theta$-functions a given term can contain -- in order not to be zero --
is two. Also, the arguments of the two $\Theta$-functions have to coincide as it is the case with the terms of class $C$. We conclude that
we only have to consider the cases $A, B$ and $C$. 

Contributions from the classes $A$ and $B$ are related to integrals over the torus,
\begin{equation}
{\int}_{\! \! \! {\cal T}} \! \! d^{d_s+1}x \, \Big( {\partial}_{r} G^T {\partial}_{r}
G^T {\partial}_{s} {\tilde G}^T {\partial}_{s} {\tilde G}^T
+ 4 {\partial}_{r} \Delta \, {\partial}_{r} G^T {\partial}_{s} {\tilde G}^T
 {\partial}_{s} {\tilde G}^T \Big) \, ,
\end{equation}
which are not singular at $d_s=2$.

Terms of class $C$, on the other hand, do lead to an ultraviolet-divergent integral. Consider, e.g., the term
\begin{equation}
\label{termClassC}
{\partial}_r \Delta(x) \, {\partial}_r \Delta(x) \, {\partial}_s G^T(-x) \, {\partial}_s G^T(-x) \, ,
\end{equation}
where we now have included the derivatives. In the limit $d_s \!\to \! 2$, the zero-temperature piece ${\partial}_r \Delta(x)$ amounts to
\begin{equation}
{\partial}_r \Delta(x) \propto \frac{x^r}{{x_4}^2} \,
\exp\Big[ - \frac{{\vec x}^2}{4 \gamma x_4} \Big] \, .
\end{equation}
The Taylor series of the function ${\partial}_s G^T(-x)$, evaluated at the origin, starts with a term linear in ${\vec x}$,
\begin{equation}
\label{Taylor}
{\partial}_s G^T(-x) \, = \, {\partial}_{\alpha s} G^T(-x)|_{x=0} \, x^{\alpha} + {\cal O}({\vec x}^3) \, .
\end{equation}
Inserting this term into Eq.(\ref{termClassC}), in the limit $d_s \!\to \! 2$, we end up with the following contribution in $K$,
\begin{equation}
\label{TaylorSingular}
K \, \propto \, \int \! \! d^2x \, dx_4 \, {\Big( \frac{{\vec x}}{{x_4}^2} \Big)}^2 \, e^{-{\vec x}^2/2 \gamma x_4} \, {\vec x}^2 \
\propto \ \int \! \! dx_4 \, \frac{1}{x_4} \, ,
\end{equation}
which is logarithmically divergent in the ultraviolet. One also readily checks that this is the only term that has to be
subtracted: The cubic Taylor term in the expansion of ${\partial}_s {\tilde G}^T$, Eq.(\ref{Taylor}), leads to a convergent contribution
to the integral $K$.

In order to isolate the ultraviolet singularity in Eq.(\ref{TaylorSingular}), we first cut out a sphere ${\cal S}$ of radius
$|{\cal S}| \leq \beta/2$ around the origin and decompose the integral involving the contributions of class $C$ as
\begin{eqnarray}
& & {\int}_{\! \! \! {\cal T}} \! \! d^{d_s+1} \! x \, {\partial}_r \Delta
 {\partial}_r \Delta \, {\partial}_s {\tilde G}^T {\partial}_s {\tilde G}^T
\nonumber \\
& & = {\int}_{\! \! \! {\cal S}} \! \! d^{d_s+1} \! x \, {\partial}_r \Delta
{\partial}_r \Delta \, {\partial}_s {\tilde G}^T {\partial}_s {\tilde G}^T
+ {\int}_{\! \! \! {{\cal T} \setminus \cal S}} \! \! d^{d_s+1} \! x \,
{\partial}_r \Delta {\partial}_r \Delta \, {\partial}_s {\tilde G}^T
{\partial}_s {\tilde G}^T \, .
\end{eqnarray}
In the limit $d_s \!\to \! 2$, the integral over the complement ${\cal T} \setminus {\cal S}$ of the sphere is not singular. The
divergence is contained in the integral over the sphere. Here we subtract the singular term (\ref{TaylorSingular}), which leads us to
\begin{eqnarray}
\label{sphereDecomp}
& & \hspace*{-1cm} {\int}_{\! \! \! {\cal S}} \! \! d^{d_s+1} \! x \, {\partial}_r \Delta(x) {\partial}_r
\Delta(x) \, {\partial}_s G^T(-x) {\partial}_s G^T(-x) \nonumber \\
& = & {\int}_{\! \! \! {\cal S}} \! \! d^{d_s+1} \! x \, {\partial}_r \Delta(x) {\partial}_r \Delta(x) \,
Q_{ss}(x) \nonumber \\
& & + {\int}_{\! \! \! {\cal S}} \! \! d^{d_s+1} \! x \, {\partial}_r \Delta(x) {\partial}_r \Delta(x) \,
{\partial}_{\alpha s} G^T(-x)|_{x=0} \, {\partial}_{\beta s} G^T(-x)|_{x=0} \, x^{\alpha} \, x^{\beta} \, ,
\end{eqnarray}
with $Q_{ss}(x)$ defined as
\begin{equation}
Q_{ss}(x) = {\partial}_s G^T(-x) {\partial}_s G^T(-x) \, - \, {\partial}_{\alpha s} G^T(-x)|_{x=0} {\partial}_{\beta s} G^T(-x)|_{x=0} \,
x^{\alpha} \, x^{\beta} \, .
\end{equation}
The first integral on the right hand side in Eq.(\ref{sphereDecomp}) is now convergent. The second integral, however, is divergent in the
ultraviolet. The last step in the isolation of this divergence consists in decomposing the singular integral as follows:
\begin{eqnarray}
& & \hspace*{-1cm} {\int}_{\! \! \! {\cal S}} \! \! d^{d_s+1} \! x \, {\partial}_r \Delta(x) \,
{\partial}_r \Delta(x)  \,
{\partial}_{\alpha s} G^T(-x)|_{x=0} \, {\partial}_{\beta s} G^T(-x)|_{x=0} \, x^{\alpha} \, x^{\beta}
\nonumber \\ 
& = & {\int}_{\! \! \! {\cal R}} \! \! d^{d_s+1} \! x \, {\partial}_r \Delta(x) \, {\partial}_r \Delta(x)  \,
{\partial}_{\alpha s} G^T(-x)|_{x=0} \, {\partial}_{\beta s} G^T(-x)|_{x=0} \, x^{\alpha} \, x^{\beta}
\nonumber \\
& & - {\int}_{\! \! \! {\cal R} \setminus {\cal S}} \! \! d^{d_s+1} x \, {\partial}_r \Delta(x) \,
{\partial}_r \Delta(x)  \, {\partial}_{\alpha s} G^T(-x)|_{x=0} \, {\partial}_{\beta s} G^T(-x)|_{x=0} \,
x^{\alpha} \, x^{\beta} \, .
\end{eqnarray}
Now the ultraviolet singularity is contained in the integral over all Euclidean space, which can be written as
\begin{eqnarray}
\label{singular}
& & {\int}_{\! \! \! {\cal R}} \! \! d^{d_s+1} \! x \, {\partial}_r \Delta(x) \, {\partial}_r \Delta(x)  \,
{\partial}_{\alpha s} G^T(-x)|_{x=0} \, {\partial}_{\beta s} G^T(-x)|_{x=0} \, x^{\alpha} \, x^{\beta}
\nonumber \\
& = & \frac{d_s(d_s+2)}{2^{3d_s+5} \pi^{\frac{3d_s}{2}} \gamma^{\frac{3d_s+4}{2}}} \ T^{d_s+2} \,
{(\mu H)}^{\frac{d_s-2}{2}} \, { \Bigg\{ \sum_{n=1}^{\infty} \,
\frac{e^{- \mu H n \beta}}{n^{\frac{d_s+2}{2}}} \Bigg\} }^2 \, \Gamma(1-\frac{d_s}{2}) \, .
\end{eqnarray}
In the limit $d_s \!\to \! 2$, the above regularized expression is divergent because the $\Gamma$-function develops a pole. Finally, we
subtract this divergent expression from the integral $K$ and define the renormalized integral ${\bar K}$ as
\begin{eqnarray}
\label{K bar}
{\bar K} & = & K - 2 {\int}_{\! \! \! {\cal R}} \! \! d^3x \, {\partial}_r \Delta(x) \, {\partial}_r \Delta(x) \,
{\partial}_{\alpha s} G^T(-x)|_{x=0} \, {\partial}_{\beta s} G^T(-x)|_{x=0} \, x^{\alpha} \, x^{\beta} \nonumber \\
 & = & {\int}_{\! \! \! {\cal T}} \! \! d^3x \, \Big( {\partial}_r G^T {\partial}_r G^T {\partial}_s {\tilde G}^T {\partial}_s {\tilde G}^T
+ 4 \, {\partial}_r \Delta \, {\partial}_r G^T {\partial}_s {\tilde G}^T {\partial}_s {\tilde G}^T \Big) \nonumber \\
& & + 2 {\int}_{\! \! \! {\cal T} \setminus {\cal S}} \! \! d^3x \, {\partial}_r \Delta \, {\partial}_r \Delta \, {\partial}_s {\tilde G}^T \,
{\partial}_s {\tilde G}^T + 2 {\int}_{\! \! \! {\cal S}} \! \! d^3x \, {\partial}_r \Delta \, {\partial}_r \Delta \, Q_{ss} \nonumber \\
& & - 2 \, {\int}_{\! \! \! {\cal R} \setminus {\cal S}} \! \! d^3 x \, {\partial}_r \Delta \, {\partial}_r \Delta \,
{\partial}_{\alpha s} G^T(-x)|_{x=0} \, {\partial}_{\beta s} G^T(-x)|_{x=0} \, x^{\alpha} \, x^{\beta} \, .
\end{eqnarray}
While the numerical evaluation of the finite expression ${\bar K}$ poses no problems and is considered in appendix \ref{AppendixA}, here
we proceed with our discussion regarding the divergent piece in $K$. The point is that this divergence originating from the three-loop
graph 8c can be canceled by an identical divergence in the two-loop graph 8d. More precisely, the divergence in the three-loop graph 8c
can be absorbed into the next-to-leading effective constants $l_1$ and $l_2$ which come along with the two-loop graph 8d. This logarithmic
renormalization of higher-order effective constants is standard in a Lorentz-invariant framework. Indeed, the absorption of ultraviolet
divergences occurring in loop graphs into next-to-leading order effective constants has been discussed a long time ago in the context of
chiral perturbation theory, i.e. the low-energy effective theory of Quantum Chromodynamics \citep{GL85}. Since the issue is rather
technical, we do not present it in the main body of the paper -- rather we relegate it to appendix \ref{AppendixB}.

Collecting all terms contributing to the free energy density up to order $p^8$, and writing the integral ${\bar K}$ as
\begin{equation}
{\bar K}(\sigma) = T^4 \, \frac{k(\sigma)}{\gamma^5} \, , \qquad \qquad \sigma = \mu H \beta = \frac{\mu H}{T} \, , \quad
\gamma = \frac{F^2}{\Sigma} \, ,
\end{equation}
the low-temperature expansion of the free energy density for the two-dimensional ideal ferromagnet finally takes the form
\begin{eqnarray}
\label{FreeCollectOrder8}
z & = & - \Sigma \mu H  \; - \; \frac{1}{4 \pi \gamma} \, T^2 \, \sum^{\infty}_{n=1} \frac{e^{- \sigma n}}{n^2}
\; - \; \frac{l_3}{\pi \Sigma {\gamma}^3} \, T^3 \, \sum^{\infty}_{n=1} \frac{e^{- \sigma n}}{n^3} \nonumber \\
& & - \frac{3( 4 l_3^2 - c_1 \Sigma \gamma)}{\pi {\Sigma}^2 {\gamma}^5} \, T^4 \, \sum^{\infty}_{n=1}
\frac{e^{- \sigma n}}{n^4} \nonumber \\
& & - \frac{8 {\overline l}_1 + 6 {\overline l}_2 + 5 l_3}{24 {\pi}^2 {\Sigma}^2 {\gamma}^4} \, T^4 \, {\Bigg\{ \sum^{\infty}_{n=1}
\frac{e^{- \sigma n}}{n^2} \Bigg\}}^2 \; -  \; \frac{1}{2{\Sigma}^2 {\gamma}^3} \, k(\sigma) \, T^4 \; + \; {\cal O}(p^{10}) \, .
\end{eqnarray}
The quantities ${\overline l}_1$ and ${\overline l}_2$ are the renormalized next-to-leading effective constants (see appendix
\ref{AppendixB}). Note that the next-to-leading effective constant $l_3$, much like the next-to-next-to-leading effective constant $c_1$,
does not require renormalization -- both quantities are finite as they stand.

While this series has been investigated before \citep{ML69,Col72,YK73,Tak86,Tak87a,Tak87b,Tak90,AA90,SSI94,NT94,Yab91,KSK03},
both within modified spin-wave theory and Schwinger-Boson mean field theory, we want to emphasize that all these references dealt with
{\it free} magnons. The manifestation of the spin-wave interaction in the low-temperature expansion for the free energy density of the
two-dimensional ideal ferromagnet is considered here for the first time to the best of our knowledge. Furthermore, our rigorous approach
is completely systematic and does not resort to any kind of approximations or ad hoc assumptions.

\section{Thermodynamics of Two-Dimensional Ideal Ferromagnets}
\label{Thermodyn}

Using the representation (\ref{FreeCollectOrder8}) for the free energy density we now discuss the thermodynamic properties of
two-dimensional ideal ferromagnets. Let us first consider the low-temperature series for the pressure. Because the system is homogeneous,
the pressure can be obtained from the temperature-dependent part of the free energy density,
\begin{equation}
\label{Pz}
P \, = \, z_0 - z \, .
\end{equation}
Accordingly, up to order $p^8$, the low-temperature series for the pressure takes the form
\begin{equation}
\label{Pressure}
P \; = \; {\eta}_0 \, T^2 \, + \, {\eta}_1 \, T^3 \, + \, {\eta}_2 \, T^4 \, + \, {\cal O}(p^{10}) \, ,
\end{equation}
where the coefficients ${\eta}_i$ are given by
\begin{eqnarray}
\label{PressureCoefficients}
{\eta}_0 & = & \frac{1}{4 \pi \gamma} \, \sum^{\infty}_{n=1} \frac{e^{- \sigma n}}{n^2} \, , \nonumber \\
{\eta}_1 & = & \frac{l_3}{{\pi} \Sigma {\gamma}^3} \, \sum^{\infty}_{n=1} \frac{e^{- \sigma n}}{n^3} \, , \nonumber \\
{\eta}_2 & = & \frac{3( 4 l_3^2 - c_1 \Sigma \gamma)}{\pi {\Sigma}^2 {\gamma}^5} \, \sum^{\infty}_{n=1} \frac{e^{- \sigma n}}{n^4}
\nonumber \\
& & + \frac{8 {\overline l}_1 + 6 {\overline l}_2 + 5 l_3}{24 {\pi}^2 {\Sigma}^2 {\gamma}^4} \, {\Bigg\{ \sum^{\infty}_{n=1}
\frac{e^{- \sigma n}}{n^2} \Bigg\}}^2 + \frac{1}{2{\Sigma}^2 {\gamma}^3} \, k(\sigma) \, .
\end{eqnarray}
Note that these coefficients depend on the dimensionless ratio $\sigma = \mu H / T$. Remarkably, the spin-wave interaction starts
manifesting itself only at order $p^8 \propto T^4$ through the last two terms in the coefficient ${\eta_2}$. While the former
contribution, which involves the renormalized effective constants ${\overline l}_1$ and ${\overline l}_2$, originates from a two-loop
graph, the latter contribution, proportional to the dimensionless function $k(\sigma)$, comes from a three-loop graph. All other
contributions in the pressure are related to one-loop graphs, i.e. to graphs that describe noninteracting magnons. In the above series for
the pressure they contribute at order $T^2, T^3$ and $T^4$.

The ratio $\sigma = \mu H / T$ in the series for the pressure can take any value, provided that the temperature and the magnetic field are
both small with respect to the intrinsic scale of the underlying theory. In the present case of the two-dimensional ideal ferromagnet,
this scale may be identified with the exchange integral $J$ of the Heisenberg model. In the following we will be interested in the limit
$T \gg \mu H$ which we implement by holding $T$ fixed and sending the magnetic field to zero. By keeping the fixed temperature small
compared to the scale $J$, we thus never leave the domain of validity of the low-temperature effective expansion.

\begin{figure}
\includegraphics[width=13cm]{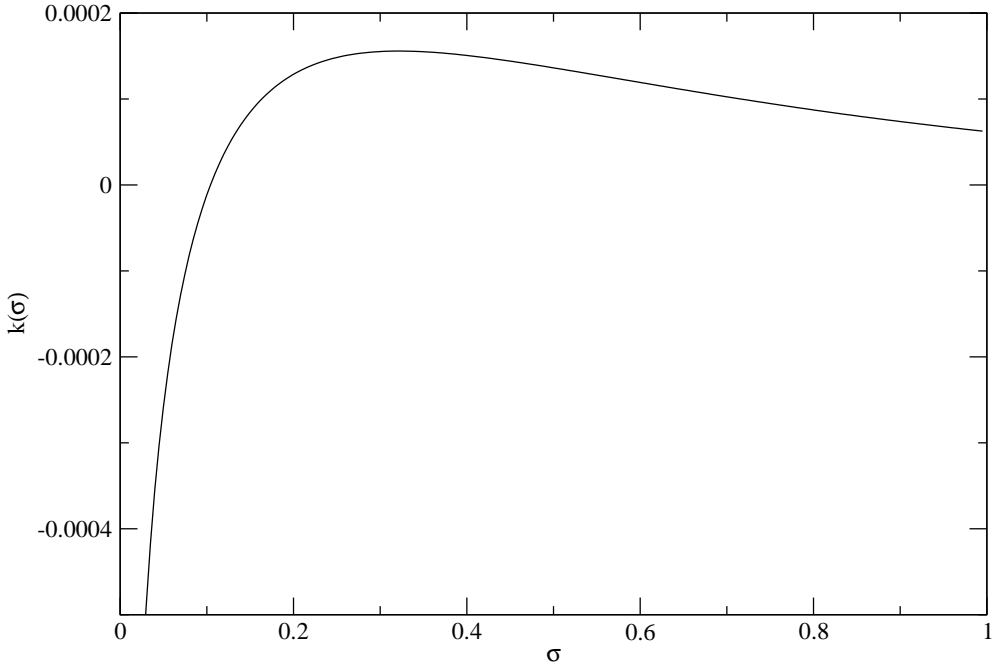}
\caption{The function $k(\sigma)$, where $\sigma$ is the dimensionless parameter $\sigma = \mu H / T$.}
\label{figure2}
\end{figure}

Formally, the limit $\sigma \to \! 0$ poses no problems for the one-loop contributions in the pressure. The corresponding coefficients in
(\ref{PressureCoefficients}) become temperature independent and the sums reduce to Riemann zeta functions. However, one has to be very
careful by taking this limit in the interaction part contained in the coefficient ${\eta}_2$. In order to address this issue, we first
take a closer look at the dimensionless function $k(\sigma)$ which is depicted in Fig.~\ref{figure2}.

In the limit $\sigma \! \to \!0$, this function can be parametrized as
\begin{equation}
\label{representationFunctionk}
k(\sigma) = k_0 \ln \sigma + k_1 +  k_2 \sigma \ln^2 \! \sigma + k_3 \, \sigma \ln \sigma + k_4 \, \sigma
+ {\cal O}(\sigma^2 \ln^3 \! \sigma) \, ,
\end{equation}
where the coefficients $k_i$ are pure numbers. Here we only need $k_0$ and $k_1$ which take the values
\begin{equation}
k_0 = \frac{\pi}{4608} = 6.818 \times 10^{-4} \, , \qquad k_1 = 2.18 \times 10^{-3} \, .
\end{equation}
The function $k(\sigma)$ thus diverges logarithmically, indicating that the limit $\sigma \! \to \!0$ in the above representation for the
pressure is rather subtle. In appendix \ref{AppendixC} it is shown that the sum of the two-loop and the three-loop contribution in the
pressure,
\begin{equation}
\frac{8 {\overline l}_1 + 6 {\overline l}_2 + 5 l_3}{24 {\pi}^2 {\Sigma}^2 {\gamma}^4} \, {\Bigg\{ \sum^{\infty}_{n=1}
\frac{e^{- \sigma n}}{n^2} \Bigg\}}^2 \, T^4 + \frac{1}{2{\Sigma}^2 {\gamma}^3} \, k(\sigma) \, T^4 \, ,
\end{equation}
remains finite in the limit $\sigma \! \to \!0$ and can be written as
\begin{equation}
\frac{k_0}{2 {\Sigma}^2 {\gamma}^3} \, T^4 \, \ln \Big( \frac{\Lambda_p}{T} \Big)
+ \frac{5 l_3}{24 {\pi}^2 {\Sigma}^2 {\gamma}^4} {\zeta(2)}^2 \, T^4 \, .
\end{equation}
The quantity $\Lambda_p$ is a scale which depends on the renormalized next-to-leading order effective constants ${\overline l}_1$ and
${\overline l}_2$, as well as on the coefficients $k_0$ and $k_1$. In the absence of a magnetic field, the low-temperature expansion for
the pressure of the two-dimensional ideal ferromagnet hence takes the form
\begin{eqnarray}
\label{PressureZeroH}
P & = & \frac{1}{4 \pi \gamma} \, \zeta(2) \, T^2 \; + \; \frac{l_3}{\pi \Sigma {\gamma}^3} \, \zeta(3) \, T^3
+ \frac{3( 4 l_3^2 - c_1 \Sigma \gamma)}{\pi {\Sigma}^2 {\gamma}^5} \, \zeta(4) \,T^4 \nonumber \\
& & + \frac{5 l_3}{24 {\pi}^2 {\Sigma}^2 {\gamma}^4} \, {\zeta(2)}^2 \, T^4
+ \frac{k_0}{2 {\Sigma}^2 {\gamma}^3} \, T^4 \, \ln \Big( \frac{{\Lambda}_p}{T} \Big) \; + \; {\cal O}(p^{10}) \, .
\end{eqnarray}
Interestingly, we are not dealing with a simple power series. At order $p^8$, where the spin-wave interaction sets in, a logarithmic
contribution $T^4 \ln T$ emerges. Although this term contains a scale related to next-to-leading order effective constants of
${\cal L}^4_{eff}$, the coefficient of this term is fixed by the effective constants $\Sigma$ and $F$ of the leading order Lagrangian
${\cal L}^2_{eff}$ and by the quantity $k_0$ which is a pure number.

In a Lorentz-invariant setting, the occurrence of such logarithmic terms is well-known. The low-temperature expansion of the pressure in
Quantum Chromodynamics involves a term $T^8 \ln T$ in the chiral limit \citep{GL89}. In fact, a term $T^8 \ln T$ also occurs in the
low-temperature expansion of the pressure of the three-dimensional antiferromagnet in the absence of a staggered field \citep{Hof99b}.
Note that antiferromagnetic magnons, much like pions in Quantum Chromodynamics, display a linear, i.e. relativistic dispersion law.

While the limit $H \to 0$ in the interaction contribution turned out to be rather subtle, one may have doubts in general about switching
off the magnetic field. After all, we are dealing with a planar system, where we are facing the Mermin-Wagner theorem. However, as shown
in Ref.~\citep{Hof12}, switching off the magnetic field in the low-temperature series for the pressure does not really numerically affect
the series. We will discuss the issue in more detail in the next section where we will consider the magnetization.

Let us now derive the low-temperature series for the energy density $u$, the entropy density $s$, and the heat capacity $c_V$ of the
two-dimensional ideal ferromagnet. They are readily worked out from the thermodynamic relations
\begin{equation}
\label{Thermodynamics}
s = \frac{{\partial}P}{{\partial}T} \, , \qquad u = Ts - P \, , \qquad 
c_V = \frac{{\partial}u}{{\partial}T} = T \, \frac{{\partial}s}{{\partial}T} \, ,
\end{equation}
and amount to
\begin{eqnarray}
u & = &  \frac{1}{4 \pi \gamma} \, T^2 \, \Bigg\{ \sum^{\infty}_{n=1} \frac{e^{- \sigma n}}{n^2}
+ \sigma \sum^{\infty}_{n=1} \frac{e^{- \sigma n}}{n} \Bigg\} \nonumber \\
& & + \frac{l_3}{\pi \Sigma {\gamma}^3} \, T^3 \, \Bigg\{ 2 \sum^{\infty}_{n=1} \frac{e^{- \sigma n}}{n^3}
+ \sigma \sum^{\infty}_{n=1} \frac{e^{- \sigma n}}{n^2} \Bigg\} \nonumber \\
& & + \frac{3( 4 l_3^2 - c_1 \Sigma \gamma)}{\pi {\Sigma}^2 {\gamma}^5} \, T^4 \, \Bigg\{ 3 \sum^{\infty}_{n=1}
\frac{e^{- \sigma n}}{n^4} + \sigma \sum^{\infty}_{n=1} \frac{e^{- \sigma n}}{n^3} \Bigg\} \nonumber \\
& & + \frac{8 {\overline l}_1 + 6 {\overline l}_2 + 5 l_3}{24 {\pi}^2 {\Sigma}^2 {\gamma}^4} \, T^4 \,
\Bigg\{ 3 \sum^{\infty}_{m=1} \frac{e^{- \sigma m}}{m^2} \sum^{\infty}_{n=1} \frac{e^{- \sigma n}}{n^2}
+ 2 \sigma \sum^{\infty}_{m=1} \frac{e^{- \sigma m}}{m} \sum^{\infty}_{n=1} \frac{e^{- \sigma n}}{n^2} \Bigg\} \nonumber \\
& & + \frac{1}{2 {\Sigma}^2 {\gamma}^3} \, T^4 \, \Bigg\{ 3 k(\sigma) -  \sigma \frac{d k(\sigma)}{d \sigma} \Bigg\} + {\cal O}(p^{10}) \, ,
\end{eqnarray}
\begin{eqnarray}
s & = &  \frac{1}{4 \pi \gamma} \, T \, \Bigg\{ 2 \sum^{\infty}_{n=1} \frac{e^{- \sigma n}}{n^2}
+ \sigma \sum^{\infty}_{n=1} \frac{e^{- \sigma n}}{n} \Bigg\} \nonumber \\
& & + \frac{l_3}{\pi \Sigma {\gamma}^3} \, T^2 \, \Bigg\{ 3 \sum^{\infty}_{n=1} \frac{e^{- \sigma n}}{n^3}
+ \sigma \sum^{\infty}_{n=1} \frac{e^{- \sigma n}}{n^2} \Bigg\} \nonumber \\
& & + \frac{3( 4 l_3^2 - c_1 \Sigma \gamma)}{\pi {\Sigma}^2 {\gamma}^5} \, T^3 \, \Bigg\{ 4 \sum^{\infty}_{n=1}
\frac{e^{- \sigma n}}{n^4} + \sigma \sum^{\infty}_{n=1} \frac{e^{- \sigma n}}{n^3} \Bigg\} \nonumber \\
& & + \frac{8 {\overline l}_1 + 6 {\overline l}_2 + 5 l_3}{12 {\pi}^2 {\Sigma}^2 {\gamma}^4} \, T^3 \,
\Bigg\{ 2 \sum^{\infty}_{m=1} \frac{e^{- \sigma m}}{m^2} \sum^{\infty}_{n=1} \frac{e^{- \sigma n}}{n^2}
+ \sigma \sum^{\infty}_{m=1} \frac{e^{- \sigma m}}{m} \sum^{\infty}_{n=1} \frac{e^{- \sigma n}}{n^2} \Bigg\} \nonumber \\
& & + \frac{1}{2 {\Sigma}^2 {\gamma}^3} \, T^3 \, \Bigg\{ 4 k(\sigma) -  \sigma \frac{d k(\sigma)}{d \sigma} \Bigg\}
+ {\cal O}(p^8) \, ,
\end{eqnarray}
\begin{eqnarray}
c_V & = &  \frac{1}{4 \pi \gamma} \, T \, \Bigg\{  2 \sum^{\infty}_{n=1} \frac{e^{- \sigma n}}{n^2}
+ 2 \sigma \sum^{\infty}_{n=1} \frac{e^{- \sigma n}}{n}
+ {\sigma}^2 \sum^{\infty}_{n=1} e^{- \sigma n} \Bigg\} \nonumber \\
& & + \frac{l_3}{\pi \Sigma {\gamma}^3} \, T^2 \, \Bigg\{ 6 \sum^{\infty}_{n=1} \frac{e^{- \sigma n}}{n^3}
+ 4 \sigma \sum^{\infty}_{n=1} \frac{e^{- \sigma n}}{n^2}
+ {\sigma}^2 \sum^{\infty}_{n=1} \frac{e^{- \sigma n}}{n} \Bigg\} \nonumber \\
& & + \frac{3( 4 l_3^2 - c_1 \Sigma \gamma)}{\pi {\Sigma}^2 {\gamma}^5} \, T^3 \, \Bigg\{ 12 \sum^{\infty}_{n=1}
\frac{e^{- \sigma n}}{n^4} + 6 \sigma \sum^{\infty}_{n=1} \frac{e^{- \sigma n}}{n^3} + {\sigma}^2 \sum^{\infty}_{n=1} \frac{e^{- \sigma n}}{n^2}
\Bigg\} \nonumber \\
& & + \frac{8 {\overline l}_1 + 6 {\overline l}_2 + 5 l_3}{12 {\pi}^2 {\Sigma}^2 {\gamma}^4} \, T^3 \,
\Bigg\{ 6 \sum^{\infty}_{m=1} \frac{e^{- \sigma m}}{m^2} \sum^{\infty}_{n=1} \frac{e^{- \sigma n}}{n^2}
+ 6 \sigma \sum^{\infty}_{m=1} \frac{e^{- \sigma m}}{m} \sum^{\infty}_{n=1} \frac{e^{- \sigma n}}{n^2} \nonumber \\
& & + {\sigma}^2 \sum^{\infty}_{m=1} \frac{e^{- \sigma m}}{m} \sum^{\infty}_{n=1} \frac{e^{- \sigma n}}{n}
+ {\sigma}^2 {(e^{\sigma}-1)}^{-1}  \sum^{\infty}_{n=1} \frac{e^{- \sigma n}}{n^2}
\Bigg\} \nonumber \\
& & + \frac{1}{2 {\Sigma}^2 {\gamma}^3} \, T^3 \, \Bigg\{ 12 k(\sigma) - 6 \sigma \frac{d k(\sigma)}{d \sigma}
+ {\sigma}^2 \frac{d^2 k(\sigma)}{d {\sigma}^2} \Bigg\} 
+ {\cal O}(p^8) \, .
\end{eqnarray}
In the above series, the spin-wave interaction manifests itself in the last two terms which originate from two-loop and three-loop graphs.
While the former involves the renormalized next-to-leading order effective constants ${\overline l}_1$ and ${\overline l}_2$, the latter
is proportional to the dimensionless function $k(\sigma)$ and its derivatives.

If we switch off the magnetic field, these low-temperature series turn into
\begin{eqnarray}
u & = &  \frac{1}{4 \pi \gamma} \, \zeta(2) \, T^2 + \frac{2 l_3}{\pi \Sigma {\gamma}^3} \, \zeta(3)  \, T^3
+ \frac{9( 4 l_3^2 - c_1 \Sigma \gamma)}{\pi {\Sigma}^2 {\gamma}^5}  \zeta(4) \, T^4 \nonumber \\
& &  + \frac{5 l_3}{8 {\pi}^2 {\Sigma}^2 {\gamma}^4} \, {\zeta(2)}^2 \, T^4
+ \frac{k_0}{2 {\Sigma}^2 {\gamma}^3} \, T^4 \, \Bigg\{ 3 \ln \Big( \frac{{\Lambda}_p}{T} \Big) -1  \Bigg\} + {\cal O}(p^{10}) \, ,
\end{eqnarray}
\begin{eqnarray}
s & = &  \frac{1}{2 \pi \gamma} \, \zeta(2) \, T + \frac{3 l_3}{\pi \Sigma {\gamma}^3} \, \zeta(3) \, T^2
+ \frac{12( 4 l_3^2 - c_1 \Sigma \gamma)}{\pi {\Sigma}^2 {\gamma}^5} \, \zeta(4) \, T^3 \nonumber \\
& &  + \frac{5 l_3}{6 {\pi}^2 {\Sigma}^2 {\gamma}^4} \, {\zeta(2)}^2 \, T^3
+ \frac{k_0}{2 {\Sigma}^2 {\gamma}^3} \, T^3 \, \Bigg\{ 4 \ln \Big( \frac{{\Lambda}_p}{T} \Big) - 1 \Bigg\} + {\cal O}(p^8) \, ,
\end{eqnarray}
\begin{eqnarray}
c_V & = &  \frac{1}{2 \pi \gamma} \, \zeta(2) \, T + \frac{6 l_3}{\pi \Sigma {\gamma}^3} \, \zeta(3) \, T^2
+ \frac{36( 4 l_3^2 - c_1 \Sigma \gamma)}{\pi {\Sigma}^2 {\gamma}^5} \, \zeta(4) \, T^3 \nonumber \\
& &  + \frac{5 l_3}{2 {\pi}^2 {\Sigma}^2 {\gamma}^4} \, {\zeta(2)}^2 \, T^3
+ \frac{k_0}{2 {\Sigma}^2 {\gamma}^3} \, T^3 \, \Bigg\{ 12 \ln \Big( \frac{{\Lambda}_p}{T} \Big) - 7 \Bigg\} + {\cal O}(p^8) \, .
\end{eqnarray}
Again, the interaction part does not lead to a simple power of the temperature in the limit $H \to 0$ -- rather, it also involves a piece
logarithmic in the temperature.

We want to stress that the structure of the various low-temperature series derived in this section is an immediate consequence of the
symmetries of the two-dimensional ideal ferromagnet. Unlike modified spin-wave theory or Schwinger-Boson mean field theory, the effective
Lagrangian technique does not resort to any ad hoc assumptions or approximations, but is completely systematic. Moreover, to the best of
our knowledge, the manifestation of the spin-wave interaction in the low-temperature series of the partition function of the
two-dimensional ideal ferromagnet has never been studied before. In particular, to address this problem with spin-wave theory or
Schwinger-Boson mean field theory up to the order considered in the present analysis, appears to be beyond reach.

At the end of this section, we like to compare the low-temperature expansions for the free energy density of the ideal ferromagnet in two
and three spatial dimensions, respectively. In the absence of an external magnetic field they exhibit the following general structure,
\begin{eqnarray}
z_{d_s=2} & = & - {\tilde \eta}_0 T^2 - {\tilde \eta}_1 T^3 - {\bf {\tilde \eta}^A_2 \, T^4 } + {\bf {\tilde \eta}^B_2 \, T^4 \, ln T}
+ {\cal O}({\bf p^{10}}) \, , \nonumber \\
z_{d_s=3} & = &  - {\tilde h}_0 T^{\frac{5}{2}} - {\tilde h_1} T^{\frac{7}{2}} - {\tilde h}_2 T^{\frac{9}{2}} - {\bf {\tilde h_3} T^5}
- {\bf {\tilde h_4} T^{11/2}} + {\cal O}({\bf p^{12}}) \, ,
\end{eqnarray}
where we have highlighted all contributions which are related to the spin-wave interaction. Note that in the case of the two-dimensional
ideal ferromagnet, the coefficient ${\tilde \eta}_2$ has two parts. The former one, ${\tilde \eta}^A_2$, contains the free magnon part as
well as the interaction contribution proportional to $l_3$. The latter one, ${\tilde \eta}^B_2$, is exclusively due to the spin-wave
interaction which involves the renormalized next-to-leading order effective constants ${\overline l}_1$ and ${\overline l}_2$.

In three dimensions, the term of order $T^5 \propto p^{10}$ in the free energy density is the famous Dyson interaction term -- in the
effective framework this is a two-loop effect (see Ref.~\citep{Hof02}). Corrections to the Dyson term were considered in
Ref.~\citep{Hof11a}, pointing out that the leading correction is of order $T^{11/2} \propto p^{11}$ and originates from a three-loop graph.
Interestingly, in the case of the three-dimensional ideal ferromagnet, the two-loop and the three-loop contributions occur at different
orders in the low-temperature expansion. As we have outlined in the present article, the same two-loop and three-loop graphs, in the case
of the two-dimensional ideal ferromagnet, all contribute at the same order $p^8$ and lead to a logarithmic term $T^4 \ln T$ in the free
energy density. As argued in  Ref.~\citep{Hof12}, the different organization of Feynman diagrams in three and two spatial dimensions is a
consequence of the suppression of ferromagnetic loops in Feynman graphs: In three spatial dimensions, each loop in a Feynman diagram is
suppressed by three powers of momentum. In two dimensions, on the other hand, ferromagnetic loops are only suppressed by two powers of
momentum.

\section{Magnetization and Mermin-Wagner Theorem}
\label{Magnetization}

Let us now consider the magnetization. With the expression for the free energy density (\ref{FreeCollectOrder8}), the low-temperature
expansion for the magnetization 
\begin{equation}
\Sigma(T,H) \, = \, - \frac{\partial z}{\partial(\mu H)}
\end{equation}
of the two-dimensional ideal ferromagnet takes the form
\begin{equation}
\frac{\Sigma(T,H)}{\Sigma} \; = \; 1 - {\hat \alpha}_0 \, T - {\hat \alpha}_1 \, T^2 - {\hat \alpha}_2 \, T^3 + {\cal O}(p^8) \, .
\end{equation}
The coefficients $\hat \alpha_i$ depend on the dimensionless ratio $\sigma = \mu H/T$ and are given by
\begin{eqnarray}
\label{SigmaCollectT(H=0)}
{\hat \alpha}_0 & = & \frac{1}{4 \pi \gamma \Sigma} \, \sum^{\infty}_{n=1} \frac{e^{- \sigma n}}{n} \, , \nonumber \\
{\hat \alpha}_1 & = & \frac{l_3}{{\pi} {\Sigma}^2 {\gamma}^3} \, \sum^{\infty}_{n=1} \frac{e^{- \sigma n}}{n^2}\, , \nonumber \\
{\hat \alpha}_2 & = & \frac{3( 4 l_3^2 - c_1 \Sigma \gamma)}{\pi {\Sigma}^3 {\gamma}^5} \, \sum^{\infty}_{n=1} \frac{e^{- \sigma n}}{n^3}
\nonumber \\
& & + \frac{8 {\overline l}_1 + 6 {\overline l}_2 + 5 l_3}{12 {\pi}^2 {\Sigma}^3 {\gamma}^4} \, \sum^{\infty}_{m=1} \frac{e^{- \sigma m}}{m}
\sum^{\infty}_{n=1} \frac{e^{- \sigma n}}{n^2} - \frac{1}{2{\Sigma}^3 {\gamma}^3} \, \frac{d k(\sigma)}{d \sigma} \, .
\end{eqnarray}
Obviously this series has a problem if we want to switch off the magnetic field, as the leading coefficient ${\hat \alpha}_0$ then
diverges. The point is that it is completely inconsistent to take the limit $\sigma \to 0$ in the low-temperature expansion of the
magnetization, as we will discuss now.

We should keep in mind that the present effective calculation is based on the assumption that the internal spin symmetry O(3) is
spontaneously broken. While this assumption is fulfilled at zero temperature, at finite temperature, however, spontaneous symmetry
breaking in the two-dimensional Heisenberg ferromagnet cannot occur according to the Mermin-Wagner theorem \citep{MW68}. Rather, at
finite temperature, the low-energy spectrum of the two-dimensional ideal ferromagnet exhibits a nonperturbatively generated energy gap and
the correlation length of the magnons, although still exponentially large \citep{KC89},
\begin{equation}
\label{npcorrelation}
\xi_{np} = C_{\xi} a S^{-\frac{1}{2}} \, \sqrt{\frac{T}{J S^2}} \, \exp \! \Big[\frac{2 \pi J S^2}{T} \Big]  \, ,
\end{equation}
no longer is infinite.

We have mentioned before that these nonperturbative effects are so tiny that they cannot manifest themselves in the low-temperature series
for the free energy density. Likewise, the series for the pressure, the energy density, the entropy density, and the heat capacity derived
in the previous section are also valid as they stand -- the subtleties raised by the Mermin-Wagner theorem are not relevant for these
thermodynamic quantities.

However, for the magnetization matters are quite different. There, the nonperturbatively generated energy gap does not lead to tiny
corrections in the low-temperature expansion -- rather these effects become the dominant ones. We can easily see this by expanding the
series in the small parameter $\sigma = \mu H/T$. While the free energy density amounts to
\begin{equation}
\label{FreeCollectSmallH}
z = - \Sigma \mu H - \frac{1}{4 \pi \gamma} \, T^2  \, \Big\{ \zeta(2) + \sigma \ln \sigma - \sigma - \frac{{\sigma}^2}{4}
+ {\cal O}({\sigma}^3) \Big\} + {\cal O}(T^3) \, ,
\end{equation}
the magnetization takes the form
\begin{equation}
\label{MagnetizationSmallH}
\Sigma(T,H) = \Sigma + \frac{1}{4 \pi \gamma} \, T  \, \Big\{ \ln \sigma - \frac{\sigma}{2} + \frac{{\sigma}^2}{24}
+ {\cal O}({\sigma}^4) \Big\} + {\cal O}(T^2) \, .
\end{equation}
In the effective theory perspective, the quantity ${\Delta E}_H$,
\begin{equation}
{\Delta E}_H = \mu H \, ,
\end{equation}
is the energy gap induced by the weak external magnetic field -- this is the quantity that appears in the parameter $\sigma$. However,
there is a further mechanism contributing to the energy gap, the one that generates a gap nonperturbatively. In analogy to the definition
of the correlation length, related to the energy gap ${\Delta E}_H$ \citep{Hof12},
\begin{equation}
\label{Hcorrelation}
\xi = \sqrt{\frac{\gamma}{\mu H}} = \sqrt{\frac{\gamma}{{\Delta E}_H}} \, ,
\end{equation}
the nonperturbatively generated energy gap ${\Delta E}_{np}$ is connected to the correlation length $\xi_{np}$ by
\begin{equation}
{\Delta E}_{np} = \frac{\gamma}{\xi^2_{np}} \, .
\end{equation}
In our low-temperature series, so far only the energy gap ${\Delta E}_H$ is accounted for in the parameter $\sigma$.

We have to remember that we implement the limit $T \gg \mu H$ by holding $T$ fixed and sending the magnetic field to zero. Keeping the
fixed temperature small compared to the underlying scale given by the exchange integral $J$, we never leave the domain of validity of the
effective low-temperature expansion. Now for a fixed value of the temperature, the nonperturbatively generated energy gap ${\Delta E}_{np}$
is just a constant. However, unlike the energy gap induced by the external magnetic field, the generation of ${\Delta E}_{np}$ is a
dynamical effect we cannot control. In particular, while we can switch off the magnetic field and thus ${\Delta E}_H$, we cannot switch off
${\Delta E}_{np}$ -- it would be inconsistent to consider values for $\sigma$ smaller than $\sigma_{np} ={\Delta E}_{np}/T$ in the
effective low-temperature series.

At very small temperatures -- let us say $T/J = 1/100$ -- the relations (\ref{npcorrelation}) and (\ref{Hcorrelation}) imply that the
condition
\begin{equation}
{\Delta E}_{np} = {\Delta E}_H \,
\end{equation}
is satisfied if the ratio $\mu H /T$ takes the value
\begin{equation}
\frac{\mu H}{T} = \frac{10^4}{16 C^2_{\xi}} \ e^{-100 \pi} \approx  10^{-131} \qquad \quad (S = \mbox{$\frac{1}{2}$}) \, .
\end{equation}
This tiny ratio thus leads to a negligible contribution to the free energy density (\ref{FreeCollectSmallH}) -- here, although
conceptually inconsistent, taking the limit $\sigma \to 0$ does not numerically affect the series. However, in the magnetization
(\ref{MagnetizationSmallH}) the effect of the nonperturbatively generated energy gap is rather large because of the logarithm
$\ln \sigma$ -- taking the limit $\sigma \to 0$ in the magnetization would be completely inconsistent.

In conclusion, at nonzero temperature the nonperturbatively generated correlation length is finite and there is always an energy gap
${\Delta E}_{np}$ in the spectrum of the two-dimensional ideal ferromagnet. The question is whether or not this gap is numerically
relevant in the low-temperature series -- for the thermodynamic quantities it is irrelevant, for the magnetization, however, it is indeed
relevant.

The low-temperature expansion of the magnetization of two-dimensional ideal ferromagnets has been considered before within the
formalism of double-time-tempera-ture Green functions and spin-wave theory. The explicit expressions given in Ref.~\citep{Yab91} and
Ref.~\citep{KSK03}, respectively, are consistent with our effective analysis provided that we express the effective constants $\gamma$ and
$l_3$ in terms of microscopic constants as \citep{Hof12}
\begin{equation}
\label{gammamicro}
\gamma = J S a^2 \, , \qquad l_3 = \frac{J S^2 a^2 }{32} \, .
\end{equation}
Here $J$ is the exchange integral of the Heisenberg model and $a$ is the distance between the sites on the square lattice.

We have to emphasize, however, that the references \citep{Yab91,KSK03} were restricted to {\it free} magnons. The effect of the
spin-wave interaction on the low-temperature expansion of the magnetization of two-dimensional ideal ferromagnets has been considered for
the first time in the present work.

\section{Conclusions}
\label{Summary}

Using the method of effective Lagrangians, we have evaluated the partition function of the two-dimensional ideal ferromagnet up to three
loops and have derived the low-temperature series for various thermodynamic quantities including the magnetization. In particular, we have
shown that in the absence of an external magnetic field, the spin-wave interaction starts manifesting itself in the form of a logarithmic
term $T^4 \, \ln T$ in the free energy density. To obtain this result, we had to renormalize and numerically evaluate a specific
three-loop graph which turned out to be the piece of resistance. Much like in the case of three-dimensional ideal ferromagnets, the
spin-wave interaction is also very weak in two-dimensional ideal ferromagnets.

We have discussed in detail the implications of the Mermin-Wagner theorem for the low-temperature series derived in the present work.
While the series for the free energy density, pressure, internal energy density, entropy density and specific heat capacity are also valid
if the magnetic field is switched off, one has to be careful in the case of the magnetization where the effect of the nonperturbatively
generated energy gap cannot be neglected.

Although various authors have considered the low-temperature properties of two-dimensional ideal ferromagnets before within spin-wave
theory, Schwinger-Boson mean field theory and double-time-temperature Green functions, to the best of our knowledge, they all restricted
themselves to free magnons. In particular, none of these authors identified logarithmic terms in the low-temperature expansion of the
partition function. The systematic effective field theory method thus proves to be more powerful than conventional condensed matter
methods where a three-loop analysis appears to be beyond reach. 

Such logarithmic terms are well-known in particle physics, i.e., in chiral perturbation theory which is the effective field theory of the
strong interaction described by Quantum Chromodynamics. They also occur in the context of the three-dimensional antiferromagnet, whose
spin waves obey a linear (relativistic) dispersion law. These logarithmic terms are a consequence of the structure of the ultraviolet
divergences in the Goldstone boson propagator. In Quantum Chromodynamics, in the case of three-dimensional antiferromagnets as well as
two-dimensional ideal ferromagnets, these divergences can be absorbed into next-to-leading order effective constants by logarithmic
renormalization. Interestingly, in the case of the three-dimensional ideal ferromagnet, the ultraviolet divergences of the propagator do
not require logarithmic renormalization \citep{Hof11a}: The low-temperature series for the thermodynamic quantities and the magnetization
simply involve integer and half-integer powers of the temperature.

The present study regarding the low-temperature properties of two-dimensional ideal ferromagnets is on the same footing as the analysis
performed in Ref.~\citep{Hof11a}, which dealt with three-dimensional ideal ferromagnets. In both cases a complete and systematic analysis
of the partition functions was given up to three-loop order. In the three-dimensional case, the effect of the spin-wave interaction on the
partition function has been considered by numerous authors before, in particular by Dyson in his monumental work \citep{Dys56}. The
effective three-loop analysis presented in Ref.~\citep{Hof11a} went one step beyond Dyson. In the case of the two-dimensional ideal
ferromagnet, however, to the best of our knowledge, the effect of the spin-wave interaction on the partition function has never been
discussed so far. The present study attempts to fill this gap in the literature.

\section*{Acknowledgments}
The author would like to thank H. Leutwyler and U.-J. Wiese for useful comments regarding the manuscript.

\begin{appendix}

\section{Numerical Evaluation of the Cateye Graph}
\label{AppendixA}

In this appendix, we consider the numerical evaluation of the three-loop graph 8c. The various terms in the relevant expression
${\bar K}$,
\begin{eqnarray}
\label{K bar appendix}
{\bar K} & = & {\int}_{\! \! \! {\cal T}} \! \! d^3x \, \Big( {\partial}_r G^T {\partial}_r G^T {\partial}_s {\tilde G}^T {\partial}_s
{\tilde G}^T + 4 \, {\partial}_r \Delta \, {\partial}_r G^T {\partial}_s {\tilde G}^T {\partial}_s {\tilde G}^T \Big) \nonumber \\
& & + 2 {\int}_{\! \! \! {\cal T} \setminus {\cal S}} \! \! d^3x \, {\partial}_r \Delta \, {\partial}_r \Delta \, {\partial}_s {\tilde G}^T \,
{\partial}_s {\tilde G}^T + 2 {\int}_{\! \! \! {\cal S}} \! \! d^3x \, {\partial}_r \Delta \, {\partial}_r \Delta \, Q_{ss} \nonumber \\
& & - 2 \, {\int}_{\! \! \! {\cal R} \setminus {\cal S}} \! \! d^3 x \, {\partial}_r \Delta \, {\partial}_r \Delta \,
{\partial}_{\alpha s} G^T(-x)|_{x=0} \, {\partial}_{\beta s} G^T(-x)|_{x=0} \, x^{\alpha} \, x^{\beta} \, ,
\end{eqnarray}
only involve the variables $r \! = \! |{\vec x}|$ and $t \! = \! x_4$, such that the integrals become in fact two-dimensional,
\begin{equation}
d^3 x = 2 \pi r dr \, dt \, .
\end{equation}
A very welcome consistency check on the numerics is provided by the fact that the result has to be independent of the radius of the sphere
${\cal S}$.

It is convenient to introduce the dimensionless integration variables $\eta$ and $\xi$,
\begin{equation}
\eta = T x_4 \, , \qquad \xi = \frac{1}{2} \sqrt{\frac{T}{\gamma}} \,
|{\vec x}| \, .
\end{equation}
In the integrals over the torus which involve quartic and triple sums -- the first two terms in Eq.~(\ref{K bar appendix}) -- we first
integrate over all two-dimensional space, ending up with one-dimensional integrals in the variable $\eta$. For the quartic sum we obtain
\begin{eqnarray}
\label{quartic}
& & {\int}_{\! \! \! {\cal T}} \! \! d^3 x \, {\partial}_r G^T(x) \,
{\partial}_r G^T(x) \, {\partial}_s G^T(-x) \, {\partial}_s 
G^T(-x) \nonumber \\
& & = \frac{1}{32 {\pi}^3 {\gamma}^5} \, T^4 \,
{\int}^{1/2}_{\! \! \! -1/2} \! d \eta \, \sum^{\infty}_ {n_1 \dots n_4 =1} \,
e^{-{\sigma}(n_1 + n_2 + n_3 + n_4)} \ {\hat Q}(\eta,n_1,n_2,n_3,n_4) \, , \nonumber \\
& & {\hat Q}(\eta,n_1,n_2,n_3,n_4) = \frac{{\Bigg( \frac{1}{\eta + n_1}
+ \frac{1}{\eta + n_2} +\frac{1}{-\eta + n_3} +\frac{1}{-\eta + n_4}
\Bigg)}^{-3}}
{{\Big( (\eta + n_1)(\eta + n_2)(-\eta + n_3)(-\eta + n_4)\Big)}^2} \, ,
\end{eqnarray}
while for the triple sum we get
\begin{eqnarray}
\label{triple}
& & {\int}_{\! \! \! {\cal T}} \! \! d^3 x \, {\partial}_r \Delta(x) \,
{\partial}_r G^T(x) \, {\partial}_s G^T(-x) \,
{\partial}_s G^T(-x) \nonumber \\
& & = \frac{1}{32 {\pi}^3 {\gamma}^5} \, T^4 \,
{\int}^{1/2}_{\! \! \! 0} \! d \eta \, \sum^{\infty}_ {n_2 \dots n_4 =1} \,
e^{-{\sigma}(n_2 + n_3 + n_4)} \ {\hat Q}(\eta,0,n_2,n_3,n_4) \, , \nonumber \\
& & {\hat Q}(\eta,0,n_2,n_3,n_4) = \frac{{\Bigg( \frac{1}{\eta} +\frac{1}{\eta + n_2}
+ \frac{1}{-\eta + n_3} +\frac{1}{-\eta+n_4} \Bigg)}^{-3}}
{{\Big( \eta(\eta + n_2)(-\eta + n_3)(-\eta+n_4)\Big)}^2} \, ,
\end{eqnarray}
where
\begin{equation}
\sigma = \frac{\mu H}{T} \, , \qquad \gamma = \frac{F^2}{\Sigma} \, . 
\end{equation}
In the case of the triple sums the integration over $\eta$ only extends over the interval $[0, \frac{1}{2}]$, due to the $\Theta$-function
contained in the zero-temperature propagator $\Delta(x)$.

Note that the quantities ${\hat Q}(\eta,n_1,n_2,n_3,n_4)$ and ${\hat Q}(\eta,0,n_2,n_3,n_4)$ depend in a rather nontrivial manner on the
summation variables. The slowest convergence for the expressions Eq.~(\ref{quartic}) and Eq.~(\ref{triple}) is observed for $\sigma=0$,
because no exponential damping occurs. The numerical summation has been performed in a "Cartesian" way as follows. We first define the
vector ${\vec N_i} = (n_1, n_2, n_2, n_4)$.  The first partial sum $S_1$ in the quartic series simply corresponds to the combination 
${\vec N_1} = (1,1,1,1)$ of indices. The second partial sum $S_2$ then contains all combinations of indices in the vector ${\vec N_2}$
with at least one index equal to two: $(2,1,1,1), \dots, (2,2,2,2)$, etc. For large values of $i$ and for $\sigma=0$, the partial sums
$S_i$ converge according to $1/{S_i}^2$. Proceeding in an analogous manner for the triple sums, the asymptotic behavior turns out to be
the same.

Expressions suitable for the numerical evaluation of the remaining three integrals of Eq.(\ref{K bar appendix}) involving double sums are
\begin{eqnarray}
& & {\int}_{\! \! \! {{\cal T} \setminus {\cal S}}} \! \! d^3 x \, {\partial}_r
\Delta(x) \, {\partial}_r \Delta(x) \, {\partial}_s G^T(-x) \,
{\partial}_s G^T(-x) \nonumber \\
& & = \frac{1}{32 {\pi}^3 {\gamma}^5} \, T^4 \,
{\int}_{\! \! \! \! 0}^{S} \! d \eta \, {\int}_{\! \! \! \! \sqrt{S^2-{\eta}^2}}^{\infty} \! d \xi \, {\xi}^5 \,
\sum^{\infty}_ {n_1, n_2 =1} \, e^{-{\sigma}(n_1 + n_2)} \ {\hat P}(\xi,\eta,n_1,n_2) \, ,
 \nonumber \\
& & {\hat P}(\xi,\eta,n_1,n_2) = \frac{e^{ -{\xi}^2 \Big( \frac{2}{\eta} + \frac{1}{-\eta + n_1}
+ \frac{1}{-\eta + n_2} \Big)}}
{{\Big\{ {\eta}^2(-\eta + n_1)(-\eta + n_2)\Big\}}^2} \, ,
\end{eqnarray}
\begin{eqnarray}
& & {\int}_{\! \! \! {\cal S}} \! \! d^3 x \, {\partial}_r \Delta(x) \, {\partial}_r \Delta(x) \,
Q_{ss}(x) \\
& & = \frac{1}{32 {\pi}^3 {\gamma}^5} \, T^4 \,
{\int}_{\! \! \! \! 0}^S \! d \eta \, {\int}_{\! \! \! \! 0}^{\sqrt{S^2-{\eta}^2}} \! d \xi \, {\xi}^5 \,
\sum^{\infty}_ {n_1, n_2 =1} \, e^{-{\sigma}(n_1 + n_2 + 2\eta)} \ {\hat Q}(\xi,\eta,n_1,n_2,\sigma) \, , \nonumber
\end{eqnarray}
with 
\begin{equation}
{\hat Q}(\xi,\eta,n_1,n_2,\sigma) = \frac{e^{ -{\xi}^2 \Big( \frac{2}{\eta} + \frac{1}{-\eta + n_1}
+ \frac{1}{-\eta + n_2} \Big)} \Big[ 
\frac{e^{2 \eta \sigma}}{ { \{ ( -\eta + n_1)(-\eta + n_2) \} }^2 }
- \frac{e^{{\xi}^2 ( \frac{1}{-\eta + n_1} + \frac{1}{-\eta + n_2} ) }}{n_1^2 n_2^2}
 \Big] }
{{\eta}^4}\, ,
\end{equation}
and finally,
\begin{eqnarray}
& & {\int}_{\! \! \! {\cal R} \setminus {\cal S}} \! \! d^3 x \, {\partial}_r \Delta(x) \,
{\partial}_r \Delta(x) \, {\partial}_{s \alpha} G^T(-x)|_{x=0} \,
x^{\alpha} \, {\partial}_{s \beta} G^T(-x)|_{x=0} \, x^{\beta}
\nonumber \\
& & = \ \frac{1}{32 {\pi}^3 {\gamma}^5} \, T^4 \,
{\int}_{\! \! \! \! S}^{\infty} \! d \eta {\int}_{\! \! \! \! 0}^{\infty} \! d \xi \, {\xi}^5 \,
\sum^{\infty}_ {n_1, n_2 =1} \, e^{-\sigma(n_1 + n_2 + 2\eta)} \ {\hat R}(\xi,\eta,n_1,n_2)
\nonumber \\
& & + \ \frac{1}{32 {\pi}^3 {\gamma}^5} \, T^4 \,
{\int}_{\! \! \! \! 0}^S \! d \eta {\int}_{\! \! \! \! \sqrt{S^2-{\eta}^2}}^{\infty} \! d \xi \, {\xi}^5 \,
\sum^{\infty}_ {n_1, n_2 =1} \, e^{-\sigma(n_1 + n_2 + 2\eta)} \ {\hat R}(\xi,\eta,n_1,n_2) \, ,
\nonumber \\
& & {\hat R}(\xi,\eta,n_1,n_2) = \frac{e^{ -2{\xi}^2/\eta}}
{{\Big\{ {\eta}^2 n_1 n_2 \Big\}}^2} \, .
\end{eqnarray}
In the above integrals the radius of the sphere has been chosen as $S=\frac{1}{2}$. For large values of $i$ and for $\sigma=0$, the
partial sums $S_i$ related to the above three expressions involving double sums converge according to $1/{S_i}^2$, i.e. the asymptotic
behavior is the same as for the triple and quartic sums.

\section{Logarithmic Renormalization of the Effective Constants $l_1$ and $l_2$}
\label{AppendixB}

In this appendix we discuss how the ultraviolet divergence in
\begin{eqnarray}
\label{sing}
& & {\int}_{\! \! \! {\cal R}} \! \! d^{d_s+1} \! x \, {\partial}_r \Delta(x) \, {\partial}_r \Delta(x)  \,
{\partial}_{\alpha s} G^T(-x)|_{x=0} \, {\partial}_{\beta s} G^T(-x)|_{x=0} \, x^{\alpha} \, x^{\beta}
\nonumber \\
& = & \frac{d_s(d_s+2)}{2^{3d_s+5} \pi^{\frac{3d_s}{2}} \gamma^{\frac{3d_s+4}{2}}} \ T^{d_s+2} \,
{(\mu H)}^{\frac{d_s-2}{2}} \, { \Bigg( \sum_{n=1}^{\infty} \,
\frac{e^{- \mu H n \beta}}{n^{\frac{d_s+2}{2}}} \Bigg) }^2 \, \Gamma(1-\frac{d_s}{2}) \, ,
\end{eqnarray}
originating from the three-loop graph 8c, can be absorbed into the next-to-leading effective constants $l_1$ and $l_2$ which
show up in the two-loop contribution $z^T_{8[de]}$,
\begin{equation}
\label{zz(8de)}
z^T_{8[de]} = - \frac{8 l_1 + 6 l_2 + 5 l_3}{24 {\pi}^2 {\Sigma}^2 {\gamma}^4} \, T^4 \, {\Bigg( \sum^{\infty}_{n=1}
\frac{e^{- \mu H n \beta}}{n^2} \Bigg)}^2 \, \, .
\end{equation}
The singularity in (\ref{sing}) is due to the $\Gamma$-function which contains a pole at $d_s=2$,
\begin{equation}
\Gamma(1 - \frac{d_s}{2}) =  -\frac{2}{d_s-2} - \gamma_E + {\cal O}(d_s-2) \, ,
\end{equation}
where $\gamma_E$ is Euler's constant.

Now the two expressions (\ref{sing}) and (\ref{zz(8de)}) have the same structure. They both involve the same infinite series and are both
proportional to four powers of the temperature. Taking into account the prefactor of the integral ${\bar K}$ from Eq.~(\ref{cateye}), they
add up to
\begin{equation}
\label{sumrenorm}
- \frac{1}{24 {\pi}^2 {\Sigma}^2 {\gamma}^3} \, \Bigg\{ \frac{8 l_1 + 6 l_2 + 5 l_3}{\gamma} 
+ \frac{3}{32 \pi}  \, \Gamma(1-\frac{d_s}{2}) \, {(\mu H)}^{\frac{d_s-2}{2}} \Bigg\} \,  T^4 \, {\Bigg( \sum^{\infty}_{n=1}
\frac{e^{- \mu H n \beta}}{n^2} \Bigg)}^2 \, \, .
\end{equation}
The pole in the $\Gamma$-function can thus be absorbed into the combination $8 l_1 + 6 l_2 + 5 l_3$ of next-to-leading order effective
constants. Note however, that $l_3$ does not require renormalization. This effective constant already occurred in the one-loop result for
the free energy density (\ref{FreeCollect}) and is perfectly finite. The combination $8 l_1 + 6 l_2 $, on the other hand, is divergent and
absorbs the pole in the $\Gamma$-function as we now show explicitly.

Introducing an arbitrary renormalization scale ${\tilde \mu}$, which should not be confused with the symbol $\mu$ denoting the magnetic
moment, the divergent quantity $\lambda$,
\begin{equation}
\lambda =  \Gamma(1 - \frac{d_s}{2}) \, {(\mu H)}^{\frac{d_s-2}{2}} \, ,
\end{equation}
can be decomposed into
\begin{equation}
\lambda =  \Gamma(1 - \frac{d_s}{2}) \, {{\tilde \mu}}^{\frac{d_s-2}{2}} \,
+ \, \Gamma(1 - \frac{d_s}{2}) \, \Big[ {(\mu H)}^{\frac{d_s-2}{2}} - {{\tilde \mu}}^{\frac{d_s-2}{2}} \Big] \, .
\end{equation}
The first term contains the singularity. Note that it does not involve the magnetic field, but depends on the arbitrary renormalization
scale ${\tilde \mu}$. The second term is not singular in the limit $d_s \to 2$, but approaches the finite value
\begin{equation}
- \ln \Big( \frac{\mu H}{\tilde \mu} \Big) \, ,
\end{equation}
such that the sum (\ref{sumrenorm}) takes the form
\begin{equation}
- \frac{1}{24 {\pi}^2 {\Sigma}^2 {\gamma}^3}
\, \Bigg\{ \frac{8 {\overline l}_1({\tilde \mu}) + 6 {\overline l}_2({\tilde \mu}) + 5 l_3}{\gamma} 
- \frac{3}{32 \pi}  \,  \ln \Big( \frac{\mu H}{\tilde \mu} \Big) \Bigg\} \,  T^4 \, {\Bigg( \sum^{\infty}_{n=1}
\frac{e^{- \mu H n \beta}}{n^2} \Bigg)}^2 \, \, .
\end{equation}
The quantities ${\overline l}_1({\tilde \mu})$ and ${\overline l}_2({\tilde \mu})$ are the renormalized, i.e., finite effective constants
which depend on the renormalization scale ${\tilde \mu}$. It is important to note, however, that the curly bracket is independent of this
scale. Moreover, if we chose the renormalization scale as
\begin{equation}
{\tilde \mu} = \mu H \equiv \frac{J}{10} \, ,
\end{equation}
then the logarithm drops out and it is understood that ${\overline l}_1({\tilde \mu})$ and ${\overline l}_2({\tilde \mu})$ are the
renormalized effective constants evaluated at this specific choice of the scale. The choice that ${\tilde \mu}$ be one tenth of the
underlying scale given by the exchange integral $J$, is motivated by comparison with chiral perturbation theory. There, in order for the
logarithm to drop out, the scale is to be identified with the pion mass $M$, which is about one order of magnitude smaller than the
underlying scale ${\Lambda}_{QCD}$.

From a more practical point of view we can say that the net effect of this whole renormalization procedure is simply this: The divergence
in (\ref{sing}) can be taken care of by rewriting the expression $z^T_{8[de]}$ in terms of the two renormalized effective constants
${\overline l}_1$ and ${\overline l}_2$ evaluated at the renormalization scale ${\tilde \mu} = \mu H$ as
\begin{equation}
\label{z(8derenorm)}
z^T_{8[de]} = - \frac{8 {\overline l}_1 + 6 {\overline l}_2 + 5 l_3}{24 {\pi}^2 {\Sigma}^2 {\gamma}^4} \, T^4 \, {\Bigg( \sum^{\infty}_{n=1}
\frac{e^{- \mu H n \beta}}{n^2} \Bigg)}^2 \, .
\end{equation}

\section{Pressure at Zero Magnetic Field}
\label{AppendixC}

In this final appendix, we discuss the low-temperature expansion for the pressure at zero magnetic field. While this limit in the one-loop
contributions formally poses no problems, switching off the magnetic field in the interaction contribution of order $p^8$,
\begin{equation}
\frac{8 {\overline l}_1 + 6 {\overline l}_2 + 5 l_3}{24 {\pi}^2 {\Sigma}^2 {\gamma}^4} \, {\Bigg\{ \sum^{\infty}_{n=1}
\frac{e^{- \sigma n}}{n^2} \Bigg\}}^2 \, T^4
+ \frac{1}{2{\Sigma}^2 {\gamma}^3} \, k(\sigma) \, T^4 \, , \qquad \sigma = \frac{\mu H}{T} \, ,
\end{equation}
is rather subtle. Using the representation (\ref{representationFunctionk}) for the dimensionless function $k(\sigma)$, the interaction
contribution -- in the limit $H \to 0$ -- amounts to
\begin{equation}
\label{chiralCombination}
\frac{k_0}{2 {\Sigma}^2 {\gamma}^3}  \, \Bigg\{ \frac{{\pi}^2 (8 {\overline l}_1 + 6 {\overline l}_2 + 5 l_3)}{432 \gamma k_0}
+ \ln \Big( \frac{\mu H}{T} \Big)  + \frac{k_1}{k_0} \Bigg\}  \, T^4 \, .
\end{equation}
The effective constants ${\overline l}_1$, ${\overline l}_2$ and $l_3$ have the same dimension as $\gamma =F^2 / \Sigma$. For reasons
which become obvious in a moment, we write the renormalized effective constants as
\begin{equation}
{\overline l}_1 \equiv \gamma \, c \, \ln { \Big( \frac{\Lambda_1}{\mu H} \Big)}^{\frac{1}{16}} , \quad 
{\overline l}_2 \equiv \gamma \, c \, \ln { \Big( \frac{\Lambda_2}{\mu H} \Big)}^{\frac{1}{12}} , \qquad
c  \equiv \frac{432 k_0}{{\pi}^2} \, ,
\end{equation}
where we have introduced the two scales $\Lambda_1$ and $\Lambda_2$ which are independent of the magnetic field. Inserting this
representation for ${\overline l}_1$ and ${\overline l}_2$ into the first term in the curly bracket of the expression
(\ref{chiralCombination}), one obtains
\begin{equation}
\label{chiralCombinationLog}
\frac{k_0}{2 {\Sigma}^2 {\gamma}^3} \, \Bigg\{ \ln \Big( \frac{\Lambda_p}{\mu H} \Big)
+ \ln \Big( \frac{\mu H}{T} \Big)  + \frac{5 {\pi}^2 l_3}{432 \gamma k_0} \Bigg\} \, T^4 \, ,
\end{equation}
where the scale $\Lambda_p$ is given by
\begin{equation}
\Lambda_p = \sqrt{\Lambda_1 \Lambda_2} \ e^{k_1/k_0} \approx 24.5 \, \sqrt{\Lambda_1 \Lambda_2} \, .
\end{equation}
The essential point is that in the sum (\ref{chiralCombinationLog}) the magnetic field drops out. We conclude that, if one switches off
the magnetic field in the low-temperature expansion for the pressure, the interaction contribution does not just manifest itself in the
form of a simple power of the temperature -- rather, it also involves a logarithmic contribution which reads
\begin{equation}
\frac{k_0}{2 {\Sigma}^2 {\gamma}^3} \, T^4 \, \ln \Big( \frac{\Lambda_p}{T} \Big) \, .
\end{equation}

\end{appendix}

\end{document}